\documentclass[fleqn,usenatbib]{mnras}

\usepackage[T1]{fontenc}

\usepackage{graphicx}	% Including figure files
\usepackage{amsmath}	% Advanced maths commands
\usepackage{amssymb}	% Extra maths symbols
\usepackage{amsbsy}
\usepackage[normalem]{ulem}

\usepackage{newtxtext,newtxmath}

\DeclareRobustCommand{\VAN}[3]{#2}
\let\VANthebibliography\thebibliography
\def\thebibliography{\DeclareRobustCommand{\VAN}[3]{##3}\VANthebibliography}

\title[Uniform Modelling of Tidal Streams]{Uniform Modelling of the Stellar Density of Thirteen Tidal Streams within the Galactic Halo}

\author[J. M. Patrick et al.]{
Jeffrey M. Patrick,$^{1}$\thanks{E-mail: jmpatric@andrew.cmu.edu}
Sergey E. Koposov,$^{2,3,4}$
Matthew G. Walker$^{1}$
\\
$^{1}$McWilliams Center for Cosmology, Department of Physics, Carnegie Mellon University, 5000 Forbes Ave, Pittsburgh, PA 15213, USA\\
$^{2}$Institute for Astronomy, University of Edinburgh, Royal Observatory, Blackford Hill, Edinburgh EH9 3HJ, UK\\
$^{3}$Institute of Astronomy, University of Cambridge, Madingley Road, Cambridge CB3 0HA, UK\\
$^{4}$Kavli Institute for Cosmology, University of Cambridge, Madingley Road, Cambridge CB3 0HA, UK
}

\date{Accepted XXX. Received YYY; in original form ZZZ}

\pubyear{2022}

\begin{document}
\label{firstpage}
\pagerange{\pageref{firstpage}--\pageref{lastpage}}
\maketitle

\begin{abstract}

We present the results of fitting a flexible stellar stream density model to a collection of thirteen streams around the Milky Way, using photometric data from DES, DECaLS, and Pan-STARRS.  We construct density maps for each stream and characterise their tracks on the sky, width, and distance modulus curves along the length of each stream.  We use these measurements to compute lengths and total luminosities of streams and identify substructures. Several streams show prominent substructures, such as stream broadening, gaps, large deviations of stream tracks and sharp changes in stream densities.  Examining the group of streams as a population, as expected we find that streams with globular cluster progenitors are typically narrower than those with dwarf galaxy progenitors, with streams around 100 pc wide showing overlap between the two populations. We also note the average luminosity of globular cluster streams is significantly lower than the typical luminosity of intact globular clusters. The likely explanation is that observed globular cluster streams preferentially come from lower luminosity and lower density clusters. The stream measurements done in a uniform manner presented here will be helpful for more detailed stream studies such as identifying candidate stream members for spectroscopic follow up and stellar stream dynamical modeling.  

\end{abstract}


%%%%%%%%%%%%%%%%%%%%%%%%%%%%%%%%%%%%%%%%%%%%%%%%%%

%%%%%%%%%%%%%%%%% BODY OF PAPER %%%%%%%%%%%%%%%%%%

\section{Introduction}
As globular clusters and dwarf galaxies are accreted by the Milky Way in the process of hierarchical galaxy formation, tidal forces slowly disrupt the accreted systems and strip stars away, forming structures known as stellar streams or tidal tails (see \citet{stream_formation_Johnston_1998} and \citet{Intro_to_Streams_Newberg_2016} for overview).  These streams are approximately aligned with the orbit of their progenitor.  Stars near the L1 and L2 Lagrange points are stripped from the progenitor and enter lower or higher energy orbits.  The resulting differences in orbital frequency between stripped stars and the progenitor produce leading and trailing tails that grow longer with time.  The shape of the host gravitational potential has a large impact on the shape and formation of stellar streams, making streams useful tracers to map the density of dark matter within the Milky Way \citep{SharperViewErkal2017, High_res_gd_1_Bonaca_2020}.  Early studies on some of the first streams were able to use streams to place constraints on the mass and flattening of the dark matter halo \citep{SGR_Fit_Law_2005}.  Later studies, using improved photometric observations and simulations, improved constraints on the shape of the potential \citep{ConstrainingGD1Koposov2010, Incorporating_streams_models_Deg_2014} and our understanding of how streams and their orbits evolve in time \citep{orbit_misalignment_sanders_2013, Tidal_Debris_Sim_Hendel_2015, StraySwingScatterErkal_2016}.  The sensitivity of stellar streams to the surrounding gravitational potential also makes streams useful in determining the mass and motion of other substructure within the Galaxy --- e.g., dark matter subhaloes or the Large Magellanic Cloud (LMC).  Interactions with smaller subhaloes disrupt a stream and produce over- or under-densities, broadening, or discontinuities in the stream track \citep{SharperViewErkal2017, High_res_gd_1_Bonaca_2020, brokenatlasaliqali2020}.  On the other hand, larger objects can significantly affect the Galactic potential and influence the orbits of multiple streams \citep{TucanaIIIErkal_2018, lmc_orphan_erkal_2019}.

The study of stellar streams began to take shape with the discovery of the Sagittarius (Sgr) dwarf galaxy by \citet{SGR_discovery_Ibata_1994}, who noted that Sgr was in the process of being tidally disrupted.  Sgr provided some of the first observational data on tidal disruption and became a reference point for simulations of dwarf galaxies interacting with the Milky Way \citep{SGR_Disruption_Johnston_1995}.  The field received a boost with the advent of large sky surveys such as the Sloan Digital Sky Survey (SDSS) in 1998 \citep{SDSS_Tech_Summary_York_2000}.  By imaging thousands of square degrees, SDSS paved the way for identifying new streams and structures.  Potential streams were slowly identified as the survey continued, with tails extending from Palomar 5 detected next \citep{Pal_5_Tidal_Discovery_Odenkirchen_2001} and additional streams such as GD-1 \citep{GD_1_discovery_Grillmair_2006}, Orphan-Chenab (OC) \citep{Orphan_Discovery_Grillmair_2006, StreamsDESShipp2018, ChenabOrphanKoposov_2019}, and others \citep{NGC_5466_Detection_Grillmair_2006, Grillmair_Styx_2009} coming in the following years.  More recent data releases from surveys such as the Dark Energy Survey (DES) have allowed rapid identification of large numbers of new streams at once \citep{StreamsDESShipp2018}.  With the arrival of the astrometric data from Gaia, it became possible to detect streams in phase-space (i.e. \citet{streamfinder_original_Malhan_2018} with their \textsc{streamfinder} algorithm). Spectroscopy is also starting to become more widely used to identify specific stream members and better constrain the chemical abundances and stellar population of streams \citep{larger_extent_Caldwell_2020, Chemical_Abundances_Seven_Streams_Ji_2020, chem_prop_twelve_streams_li2021s5, brokenatlasaliqali2020}.

Recent wide area surveys have resulted in a large number of new stellar streams being discovered.  Prior work on these streams has primarily focused on one of two investigative paths.  Discoveries, such as those reported by \citet{Atlas_discovery_Koposov_2014, Panstarrs_Halo_Substructures_Bernard_2016, StreamsDESShipp2018}, often identify multiple new streams and provide some identifying information such as orbital pole location, physical characteristics such as length, width, and distance, and stellar population characteristics such as age and metallicity.  By using a single procedure for identifying and measuring the streams, the quality of the results are consistent stream to stream and stream properties can be compared easily.  Follow up studies, like \citet{SharperViewErkal2017, GD_1_relic_Li2018a, ChenabOrphanKoposov_2019, brokenatlasaliqali2020, delve_jet_ferguson2021delveing}, usually focus on single streams to develop a more complete picture, such as the specific stream track, width and density variations, distance gradients, and evolution.  However, these usually rely on an analysis tailored to the stream being studied, and specific methods can vary greatly between different papers.

In this paper we took a hybrid approach to study a group of stellar streams.  We created a generalized model that we could fit to a variety of streams using a common methodology and minimal changes to the model.  Our model uses cubic splines along the length of the stream for each parameter to still produce detailed density maps and identify characteristics such as stream tracks and distance gradients.  We fit this model to thirteen streams that lie within DES, the Dark Energy Camera Legacy Survey (DECaLS), or the Panoramic Survey Telescope and Rapid Response System (Pan-STARRS).  In addition to the stream parameters recovered by fitting our model, we also  calculate the masses and luminosities of the streams.

We organize this paper in the following way.  We describe the data we use from the surveys in Section \ref{sec:datasets}.  In Section \ref{sec:bg_models} we construct our model for foreground and background stars.  In Section \ref{sec:stage_1_fits} we describe our procedure for fitting key stream parameters using an approximate model of the streams.   We use those parameters in Section \ref{sec:stage_2_splines}, where we build a flexible and detailed model of the streams' density distributions.  We fit the model to each stream and follow up with calculating additional derived characteristics in Section \ref{sec:analysis}.  We present our results for each stream in Section \ref{sec:results}.  In Section \ref{sec:discussion} we discuss common features seen in multiple streams and connections between streams and intact globular clusters and dwarf galaxies, before concluding with Section \ref{sec:conclusion}.

\section{Datasets}
\label{sec:datasets}
This paper is based on the data from three large photometric surveys, Pan-STARRS, DES, and DECaLS \citep{PANSTARRSSurveyChambers2019, DESDR1Abbott_2018, DECALSDESIDey2019}. In this section we describe what data we selected from the surveys and the photometric systems we used.  We use the source positions in addition to two photometric passbands from each survey.  For all surveys we used the $g$ passband and prioritized using the $i$ passband as $g-i$ colour has larger dynamic range for typical stellar populations.  If the $i$ band is not available for a given dataset, we instead use the $r$ band for its improved depth compared to the $z$ band.

For Pan-STARRS, we used Data Release 1 (DR1) from Pan-STARRS Telescope \#1 (PS1) by \citet{PANSTARRSSurveyChambers2019}.  The database structure is detailed in \citet{PANSTARRSDataFlewelling2019}, and the data is accessible at the PS1 website\footnote{\url{https://panstarrs.stsci.edu/}}.  We used the PSF magnitudes from the \texttt{StackObjectThin} table, and created two columns \texttt{ra} and \texttt{dec} that merge the \texttt{(rgizy)ra} and \texttt{(rgizy)dec} columns respectively.  We remove duplicates by checking \texttt{(gi)infoflag3} to select objects in the primary stack.  We enforce the condition

\begin{multline}
    \texttt{(gi)infoflag3} \; \text{\&} \\
    \texttt{panstarrs\_dr1.detectionflags3('STACK\_PRIMARY'))} > 0.
	\label{eq:PANSTARRSstackprimary}
\end{multline}

Additionally, we separate and select stars from galaxies using the PSF-Kron method outlined in the PS1 documentation\footnote{\url{https://outerspace.stsci.edu/display/PANSTARRS/\\How+to+separate+stars+and+galaxies}}:

\begin{equation}
    \texttt{(gi)PSFMag} - \texttt{(gi)KronMag} < 0.05
	\label{eq:PANSTARRSstargalaxy}
\end{equation}

We apply this cut in both the $g$ and $i$ magnitude bands.  We select objects brighter than 22.25 mag in both bands to reduce galaxy contamination that occurs at fainter magnitudes.

For DES, we used the main table \texttt{DR1\_MAIN} from the Data Release 1 (DR1)\footnote{DES Data Release 2 (DR2) is now available, but we use DR1 as it was the most recent release when developing our program.} \citep{DESDR1Abbott_2018}.  We use the dereddened PSF magnitudes \texttt{WAVG\_MAG\_PSF\_(GRIZY)\_DERED} \citep{SFD98Schlegel1998} for each object.  Well behaved objects are selected using \texttt{FLAGS\_(GI)} < 4 \citep{DESDR1Abbott_2018}.  In addition to only using objects brighter than 23.25 mag in both bands, we followed the procedure outlined by \citet{DESStarGalaxyBeastsKoposov2015} for star-galaxy separation, requiring

\begin{equation}
    |\texttt{SPREAD\_MODEL}| < 0.003 + \texttt{SPREADERR\_MODEL}
	\label{eq:DESstargalaxy}
\end{equation}

DECaLS is one of three surveys that are part of the DESI Legacy Imaging Surveys, using the Dark Energy Camera at Cerro Tololo Inter-American Observatory \citep{DECALSDESIDey2019}.  We use the DECaLS data from Data Release 8 (DR8)\footnote{DECaLS Data Release 9 (DR9) is now available, but we use DR8 as it was the most recent release when the majority of this study was conducted} of the DESI Legacy Imaging Surveys\footnote{Column information is available at \url{https://www.legacysurvey.org/dr8/catalogs/}}. We used sources brighter than 23.25 mag in both bands and objects with \texttt{TYPE} = 'PSF' to select stars.  We removed duplicates by using the \texttt{brick\_primary} flag and by using release 8000 (8001) for sources below (above) declination 32.375 due to their overlap.  Sources near bright objects were removed using the \texttt{brightblobs} flag.  DR8 contains the DECam flux values for each source, which are converted into magnitudes in their respective bands.

For each survey we also correct magnitudes  for dust extinction.  DES already contains dereddened magnitudes, so we use that data from the survey.  For Pan-STARRS, we use the dust maps from \citet{SFD98Schlegel1998} to find E(B-V) along with $A_g/E(B-V) = 3.172$ and $A_i/E(B-V) = 1.682$ from \citet{SFDExtinctionSchlafly2011} using $R_v = 3.1$.  The DECaLS survey provides E(B-V) values from \citet{SFD98Schlegel1998} for its sources, and we use $A_g/E(B-V) = 3.214$ and $A_r/E(B-V) = 2.165$\footnote{See  \url{https://www.legacysurvey.org/dr8/catalogs/\#galactic-extinction-coefficients}}.  

Three surveys that we analyse in this paper, DES, DECaLS, and Pan-STARRS overlap significantly in some areas, with most streams covered by more than one survey footprint.  We prioritize depth when deciding between datasets for the improved sensitivity to more distant streams in addition to the overall detection of more stream stars.  Because we only reach a depth of 22.5 mag with Pan-STARRS, we only use it for Ophiuchus, which does not lie within the DES or DECaLS footprint.  For the remaining streams we fit, we prioritize DES since we can use both the $g$ and $i$ bands, while we would need to use $g$ and $r$ for DECaLS.

We mask some sources within the surveys that would potentially interfere with our stream models.  We manually mask known small structures (< 2 degrees across) with high stellar densities using circular apertures.  We use \citet{GC_catalog_Harris1996, dwarf_catalogue_mcconnachie_2012} to verify these structures are known dwarf galaxies or globular clusters and are not part of a stream.  We also mask up to 2 degrees from the edge of each survey to create a clearly defined boundary for the survey.  This boundary also ensures the region has been completely observed and contains no holes.

\section{Background Model}
\label{sec:bg_models}

In this paper, we construct stellar density models for each stellar stream.  To do this, we first need to know the distribution of non-stream stars and create a corresponding background density model.  We use this background model to describe large scale variations in background and foreground stars across the region in addition to their distribution in colour-magnitude (CMD) space.  We construct our model in the coordinate frame of $\psi_1$, $\psi_2$ which will be a coordinate system usually broadly aligned with the stream, but maybe not exactly (thus we do not use commonly adopted $\phi_1$, $\phi_2$ notation).  We assume the complete background model describing the distribution in position, colour (c), and magnitude (m), $P_{bg}(\psi_1, \psi_2, c_j, m_j)$, can be approximately factorized into a spatial ($P_{bg}(\psi_1, \psi_2)$) and CMD ($P_{bg}(c_j,m_j)$) distribution.  We first construct a model for the spatial probability distribution $P_{bg}(\psi_1, \psi_2)$, following a similar method to that used by \citet{SharperViewErkal2017}:

\begin{multline}
    P_{bg}(\psi_1,\psi_2) = M_{\psi_{2,min}}(\psi_1)+(\psi_2-\psi_{2,min})S(\psi_1) \\
    + (\psi_2-\psi_{2,min})^{2}Q(\psi_1)
	\label{eq:bg_prob_model}
\end{multline}

We use cubic splines for the average background (M) and background slope (S), and we add an additional quadratic term (Q).

Next, we divide the colour-magnitude range into pixels (with the $j$'th pixel centered at colour $c_j$ and magnitude $m_j$) to model the background CMD distribution as a discrete probability distribution of a star belonging to a certain CMD pixel, $P_{bg}(c_j,m_j)$.  We use pixels 0.02 mag wide in colour and 0.1 mag wide in magnitude and limit the colour and magnitude range of our model to $0.1 \leq g-i \leq 0.9$ and $17.5 \leq i \leq 22.5$ for Pan-STARRS, $0.1 \leq g-i \leq 0.9$ and $17.5 \leq i \leq 23.5$ for DES, and $-0.1 \leq g-r \leq 0.9$ and $17.5 \leq r \leq 22.5$ for DECaLS.  We construct $P_{bg}(c_j,m_j)$ directly from the data, grouping the background stars into their respective pixels in the colour-magnitude range.  We normalise the resulting histogram to produce $P_{bg}(c_j,m_j)$.  Finally, we construct the complete background model $P_{bg}(\psi_1, \psi_2, c_j, m_j) = P_{bg}(\psi_1, \psi_2) P_{bg}(c_j,m_j))$.

\section{Initial Stream Fitting}
\label{sec:stage_1_fits}

In this section we describe our initial procedure that we use for all the streams to fit key parameters needed for the more detailed spline model described in Section  \ref{sec:stage_2_splines}.  For each stream we select stars from their corresponding dataset following Section \ref{sec:datasets}.  We used the most recent literature available on each stream at the time they were fit to identify the largest spatial extent a stream potentially spans.  We limit the spatial range of the selected stars to an area centered on the stream and that is approximately 2 to 3 times larger than that area occupied by the stream.  The majority of streams use RA and Dec as $\psi_1$ and $\psi_2$, while for GD-1, Jhelum, and Palomar 5 we use the stream aligned coordinate systems defined by \citet{ConstrainingGD1Koposov2010, JhelumMultipleBonaca_2019, SharperViewErkal2017}, respectively. 

\subsection{Simulated Stream Colour-Magnitude Diagrams}
\label{sec:simCMDs}

The stream models that we describe in this paper rely on constructing the probability distribution of stream stars in both position and colour-magnitude space. We start by describing how we construct a simulated colour-magnitude distribution ($P_{sim}(c_j, m_j | age, [Fe/H], D)$) for a given age, metallicity ([Fe/H]), and distance modulus (D).  First, for a given age and metallicity, we create a mock stellar population by sampling the initial mass function (IMF) using the log-normal IMF by \citet{IMFChabrier2001} for $M\leq1M_\odot$ and the IMF by \citet{IMFSalpeter1955} for $M>1M_\odot$.  We use $0.09 M_\odot$ as our lower mass limit when sampling the IMF, within the $0.07-0.09 M_\odot$ range for the minimum hydrogen burning mass \citep{hbmm_burrows_2001, dynamical_masses_indi_b_c_dieterich_2018}.  We then transform the sampled masses from the IMF into magnitudes for each filter using PARSEC isochrones \citep{PARSECisochronesBressan2012}.  We use this mock stellar population to construct $P_{sim}(c_j, m_j | age, [Fe/H], D)$ by taking a mock observation at a given distance modulus.  To make this mock observation, we calculate the apparent magnitudes of stars and add to it Gaussian noise with zero mean and standard deviation $\sigma_E$ to mimic observational and systematic errors.  For a given dataset/survey (DS) and passband (B), $\sigma_E(m | DS, B) = O_{DS, B}(m) + S_{DS}$ where $O_{DS, B}(m)$ is a function based on the photometric error for a given passband and $S_{DS}$ is the approximate systematic error associated with the dataset.

We calculate the function $O_{DS, B}(m)$ by sampling stars from a 10 degree by 10 degree reference region of each dataset.  We selected regions away from the edges of the datasets or the galactic plane, and that contained a large number of stars for our entire magnitude range.  Because DES and DECaLS use a shared photometric system, we use a single reference region from DES for both of them.  We divide the stars in this region into magnitude bins and compute the median photometric error, linearly interpolating between the center of each bin to create $O_{DS, B}(m)$.  We use bins 0.25 mag wide for DES and DECaLS and 0.5 mag wide for Pan-STARRS.

\begin{figure}
    \centering
    \includegraphics[width=\columnwidth]{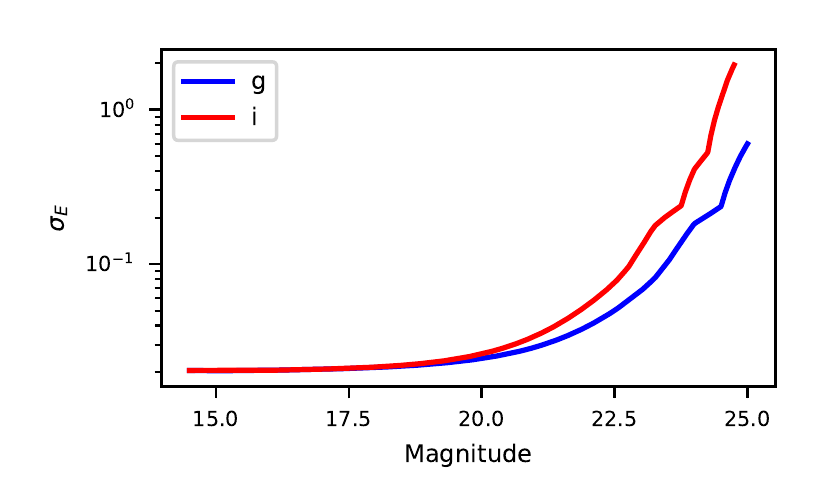}
    \caption{Assumed photometric uncertainty model ($\sigma_E$) as a function of DES $g$ and $i$ band magnitude.}
    \label{fig:des_mag_error}
\end{figure}

We set $S_{DS}$ for each dataset based on a conservative estimate for the systematic error across the whole survey footprint.  We set $S_{DS}$ to 0.01 mag for Pan-STARRS \citep{PANSTARRSErrorSchlafly2012}, 0.01 mag for DECaLS \citep{DECALSDESIDey2019}, and 0.02 mag for DES \citep{DESErrorBurke2017}.  We plot sample $\sigma_E$ curves for the $g$ and $i$ bands from DES in Figure \ref{fig:des_mag_error}.

\begin{figure}
    \centering
    \includegraphics[width=\columnwidth]{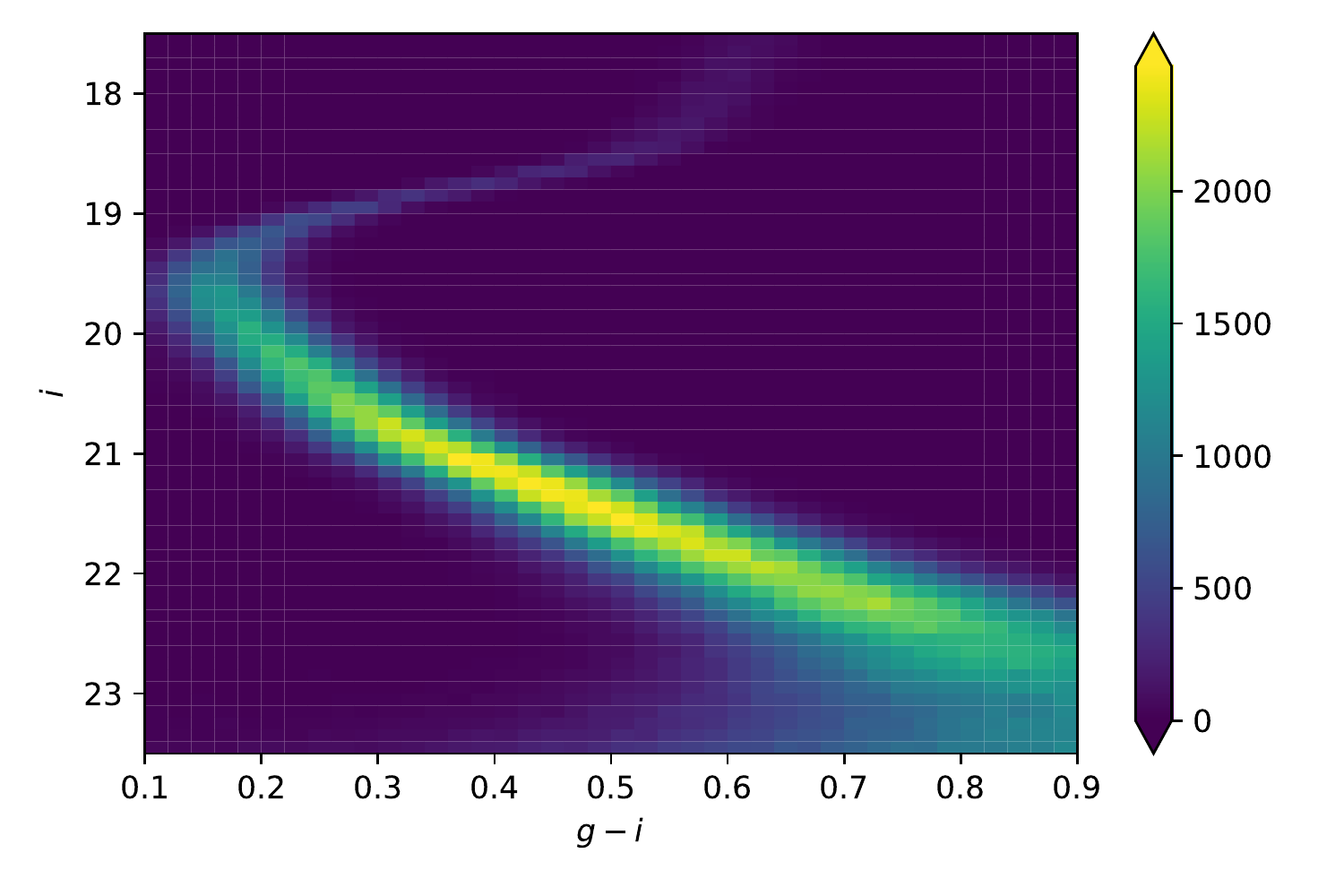}
    \caption{Histogram: Model colour magnitude distribution of 5,000,000 sampled stars from the 10 Gyr and [Fe/H]=-2.0 stellar population shifted to a distance modulus of 16.0. The photometry of simulated stars included artificially added noise. The distribution shown here is used to construct $P(c_j, m_j | age, [Fe/H], D)$.}
    \label{fig:des_scattered_CMD}
\end{figure}

After making mock observations of colours and magnitudes as described above, we then group the simulated stars using the same colour-magnitude pixels as Section \ref{sec:bg_models} (Figure \ref{fig:des_scattered_CMD}) and similarly normalise to produce $P_{sim}(c_j, m_j | age, [Fe/H], D)$.  We calculate $P_{sim}(c_j, m_j | age, [Fe/H], D)$ on a grid of ages, metallicities [Fe/H], and distance moduli and save it to avoid computationally expensive recalculations. We use a $log_{10}(age)$ range from 8.0 to 10.1 in steps of 0.05, a [Fe/H] range of $-2.1$ to 0.4 in steps of 0.1 dex, and steps of 0.02 mag in distance modulus.

\subsection{CMD Fitting}
\label{sec:cmdfitting}

With a model of the stream's probability distribution in colour-magnitude space from Section \ref{sec:simCMDs}, we are able to fit the age, metallicity ([Fe/H]), and distance modulus (D) of each stream.  We first construct a model of the expected number of observed stars in each colour-magnitude pixel, $CMD_{E}(c_j, m_j)$, in the area containing the stream.  Using an initial pole to define the stream track and initial width, $\sigma$, we define the "on stream" area as the portion of the region $<1.5\sigma$ from the center of the great circle stream track.  We create our spatial and CMD background models following the procedure set out in Section \ref{sec:bg_models}.  We use the portion of the region $>1.5\sigma$ from the center of the great circle stream track to fit the spatial background model and $>2.5\sigma$ from the stream track to fit the CMD background model.  By using different selections when fitting the spatial and CMD background models, we ensure that $P_{bg}(\psi_1, \psi_2)$ is sufficiently accurate when extrapolating across the "on stream" area while also making sure that $P_{bg}(c,m)$ contains only background stars.  We can then calculate $CMD_{E}(c,m)$ for a given number of stream stars within the on stream region (A), age, [Fe/H], and distance modulus:

\begin{multline}
    CMD_{E}(c_j, m_j | A, age, [Fe/H], D) = \\
        A \, P_{sim}(c_j, m_j | age, [Fe/H], D) + \\
        B_{>2.5\sigma} \, P_{bg}(c_j,m_j) \frac{\int_{<1.5\sigma}{P_{bg}(\psi_1, \psi_2)}}{\int_{>2.5\sigma}{P_{bg}(\psi_1, \psi_2)}}
	\label{eq:cmd_model}
\end{multline}

using $P_{sim}(c_j, m_j | age, [Fe/H], D)$ from Section \ref{sec:simCMDs}, $P_{bg}(c_j, m_j)$, and the actual number of background stars in the background section ($B_{>2.5\sigma}$) scaled using $P_{bg}(\psi_1, \psi_2)$ across the background and on stream areas.

We then used the stars within the on stream area to create $CMD_{O}(c_j, m_j)$, the actual number of observed stars within each colour-magnitude pixel.  For a given A, age, [Fe/H], and D, we can compare the estimated number of stars within a pixel to the observed number and calculate the log likelihood using Poisson statistics:

\begin{multline}
    log L(c_j,m_j)  \\
        = \log\left(\frac{CMD_{E}(c_j,m_j)^{CMD_{O}(c_j,m_j)}  e^{-CMD_{E}(c_j,m_j)}}{CMD_{O}(c_j,m_j)!}\right) \\
        \cong CMD_{O}(c_j,m_j)\log\left(CMD_{E}(c_j,m_j)\right) - CMD_{E}(c_j,m_j)
	\label{eq:cmd_negloglike}
\end{multline}

We can then sum over all CMD pixels to find the total log likelihood for that model.  Because $P_{sim}(c_j, m_j | age, [Fe/H], D)$ uses a grid of age, [Fe/H], and distance moduli, we identify the nearest 8 grid points and use multivariate linear interpolation on their total log likelihood to calculate intermediate values.  To find the maximum likelihood model over A, age, [Fe/H], and D, we first use the dynamic nested sampling \citep{DynamicNestedSamplingHigson_2018} package Dynesty \citep{NestedSamplingSkilling2004, NestedSamplingSkilling2006, BoundingMethodFeroz_2009, DynestyCodeSpeagle_2020} due to the multi-modal nature of that grid and multiple local minima possible.  We then use the Nelder-Mead algorithm to narrow down to the individual grid point with the maximum log likelihood.

It should be noted that this method underestimates the age and overestimates the metallicity and distance modulus of the stream.  All the fitted streams lie near maximum age and minimum metallicity of our simulated CMD's, resulting in truncated posterior distributions after sampling and shifting the median values lower and higher, respectively.  This is most pronounced with metallicity due to its large relative uncertainties (often greater than 0.5 dex).  Due to the degeneracy between age, metallicity, and distance modulus \citep{stellar_evolution_models_gallart_2005, bayesian_isochrone_valls_2014, estimating_ages_metallicities_howes_2019}, there will be a small increase in fit distance modulus values as well.

\subsection{Ideal Stream Matched Filter}
\label{sec:matched_filter}

With values for age, metallicity, and distance to the stream, we then construct a matched filter to select stream stars.  Specifically, we want a colour-magnitude selection that preferentially selects stars that are members of the stream while excluding background stars.  This matched filter can be constructed by assigning weights based on the colour-magnitude of the star \citep{matched_filter_weighting_Rockosi_2002} or by constructing an optimal mask in CMD space \citep{SharperViewErkal2017}. Here we adopt the second approach, selecting which colour-magnitude pixels to mask and exclude stars within them.  We construct the mask for each stream in the following way.  Using $CMD_{E}$, the expected observed CMD from the previous section, and $CMD_{str} = A \, P_{sim}(c_j, m_j | age, [Fe/H], D)$, the stream stars component of $CMD_{E}$, we look at the ratio $R = CMD_{str} / CMD_{E}$ for each CMD pixel and include pixels with a ratio larger than some threshold value, T, while masking the rest.  For each threshold value, we can calculate the expected signal-to-noise ratio:

\begin{equation}
    S/N = \left( \sum_{R \geq T} CMD_{str} \right) \bigg/ \sqrt{\sum_{R \geq T} CMD_{E}}
	\label{eq:matched_filter_s2n}
\end{equation}

We use the threshold value that maximizes the S/N ratio to create the matched filter.

\subsection{Spatial Fitting}
\label{sec:spatialfitting}

To provide a starting point for a more detailed model that we will introduce later, we use a simple stream model where it can be approximated by a great circle with the pole at $\mathbf{C} = (C_{\psi_1}, C_{\psi_2})$ and a Gaussian cross-section with a constant width $\sigma$:

\begin{equation}
    P_{str}(\psi_1, \psi_2|\mathbf{C}, \sigma) = A \, e^{({-1}/{2\sigma}) arcsin^2(\mathbf{C} \cdot \boldsymbol{\psi})}
	\label{eq:spatial_model_stream}
\end{equation}

where A is a normalising constant.  We combine $P_{str}(\psi_1, \psi_2|\mathbf{C}, \sigma)$ with the background spatial model to produce a model of all stars in the region:

\begin{multline}
    P(\psi_1, \psi_2|\mathbf{C}, \sigma, \alpha) = \\
    \alpha \, P_{str}(\psi_1, \psi_2|\mathbf{C}, \sigma) + (1-\alpha) \, P_{bg}(\psi_1, \psi_2)
	\label{eq:spatial_model}
\end{multline}

where $\alpha$ is the fraction of stars located within the region that belong to the stream.

To fit our model to the stream, we only select stars that are within our matched filter from Section \ref{sec:matched_filter}.  We fit the background model, Equation \ref{eq:bg_prob_model}, to the stars outside $>1.5\sigma$ from the starting stream track.  We use the Nelder-Mead algorithm to find the best fit values of the stream pole $\mathbf{C}$,  width $\sigma$, and $\alpha$.

After we fit the pole and width of the stream once, we repeat the CMD fit from Section \ref{sec:cmdfitting} using best fit pole and width as the new inputs.  Repeating with these new values allows us to improve the CMD fit with the updated stream position.  Similarly, after the improved CMD fit, we calculate the corresponding matched filter and repeat the spatial fit with this new matched filter.  Because not all streams might not have the same quality of initial input values, repeating each of the CMD and spatial fits once helps to provide a better starting point for the cubic spline model described in later sections.

\subsection{Stream Aligned Coordinate System}
\label{sec:coordsys}

While we have been modeling the majority of streams using RA and Dec for the initial parameter fitting, it is often more convenient to model a stream in a coordinate system aligned with the stream's track.  Therefore, we define a new coordinate system for each stream that is aligned with it, using the new coordinates $\phi_1$ and $\phi_2$ for the longitude and latitude coordinates respectively.  For most streams, we use the great circle from the pole of the spatial model as the stream track, and rotate so that the $\phi_1$ axis is aligned with it.  However, in addition to using the previously mentioned rotation matrices for GD-1, Jhelum, and Palomar 5, we use the rotation matrix from \citet{brokenatlasaliqali2020} for the ATLAS half of the AAU stream.  This allows a better comparison between our results and other papers that also use those rotation matrices.  We include the rotation matrices from RA and Dec to $\phi_1$ and $\phi_2$ for all streams in Appendix \ref{sec:rot_matrix_appendix}.

\section{Fitting the Cubic Spline Model}
\label{sec:stage_2_splines}

The stream models described in Sections \ref{sec:cmdfitting} and \ref{sec:spatialfitting} were useful to fit several stream parameters, but they used limited slices of the available data on a stream and were limited in complexity.  In order to create more complete stream models, we adapt the flexible model that \citet{SharperViewErkal2017} used to fit the spatial distribution of Palomar 5's tidal tails, into a more general model that can fit any stream using positional and colour-magnitude data.  We describe our specific model in Subsection \ref{sec:cubic_spline_model}. 

We take advantage of the initial stream fits described in Section \ref{sec:stage_1_fits} in two ways during this stage.  First, we fix the values for age and metallicity of the stream using the best-fit results from Section \ref{sec:cmdfitting}.  We assume a given stellar stream originated from a single progenitor and therefore has a consistent age and metallicity through its stellar population.  By only fitting the distance modulus to the stream, we also avoid the age-metallicity-distance degeneracy \citep{stellar_evolution_models_gallart_2005, bayesian_isochrone_valls_2014, estimating_ages_metallicities_howes_2019} that would otherwise interfere with the fit.  This allows us to model the distance modulus as a function along the length of the stream and detect distance gradients.  Second, as in our preliminary spatial fit, we use stars within our matched filter CMD mask to remove excess background stars and improve the stream contrast.

We remark that while we assume that the background model  $P_{bg}(\phi_1, \phi_2, c_j, m_j)$ can be factorized as a product of the spatial and colour-magnitude distributions, in reality this may not be the case.  Large background or foreground structures, such as the Sagittarius stream, may result in the background CMD distribution changing depending on sky position thus breaking the factorization assumption.  By using stars within the matched filter mask the impact of this effect is reduced.  We make a filter similar to Section \ref{sec:spatialfitting}, modifying it slightly to account for a possible distance gradient along the stream.  We first make a matched filter following the procedure described in Section \ref{sec:matched_filter}.  We then increase the extent of the matched filter in magnitude by expanding it by $\pm max(0.3 + 0.01 L, 0.5)$ mag above and below the filter, where L is the length of the region in $\phi_1$ in degrees.  By expanding the region, we ensure that the even if the stream has a distance gradient the stars from the stream will still be included in the fit.

\subsection{Cubic Spline Stream Model}
\label{sec:cubic_spline_model}

In this section, we describe how we expand the stream model from \citet{SharperViewErkal2017} to include colour-magnitude data on each star and fit the distance modulus gradient of the stream.  Similar to their model, we model the $\phi_2$ position ($Pos_{\phi_2}$), width ($\sigma$, the standard deviation of the Gaussian), and linear density (I) as functions of $\phi_1$ using cubic splines and use them to construct the spatial model of the stream:

\begin{multline}
    P_{spatial}(\phi_1,\phi_2 | Pos_{\phi_2}, \sigma, I) = A \, I(\phi_1 ) \frac{1}{\sqrt{2\upi} \sigma(\phi_1 )} \\
    e^{-\frac{1}{2\sigma(\phi_1)^2}(Pos_{\phi_2}(\phi_1)-\phi_2)^2}
	\label{eq:spline_stream_model_spatial}
\end{multline}

with normalisation constant A.  To calculate A, we approximate the integral of Equation \ref{eq:spline_stream_model_spatial} by sampling the region with HEALPix\footnote{\url{https://healpix.sourceforge.io/}} pixels using the healpy package \citep{healpytwo2005ApJ...622..759G, healpyoneZonca2019}.   We increase NSIDE automatically to ensure that the pixel separation is always smaller than 1.5 times minimum stream width, but cap NSIDE to 4096 for computational reasons.  We include colour-magnitude data in our model by incorporating the simulated  stream probability distributions in colour and magnitude from Section \ref{sec:simCMDs} to create a complete model of the stream in the position, colour, magnitude space:
\begin{multline}
    P_{str} (\phi_1,\phi_2,c,m | Pos_{\phi_2}, \sigma, I, D) = \\
    P_{spatial}(\phi_1,\phi_2 | Pos_{\phi_2}, \sigma, I) \, P_{sim}(c,m,\phi_1 | age, [Fe/H], D)
	\label{eq:spline_stream_model}
\end{multline}

We include an additional cubic spline for the distance modulus along the length of the stream ($D(\phi_1)$).  Because $D(\phi_1)$ is continuous and we generate $P_{sim}$ using a grid of distance moduli, we linearly interpolate the individual pixel values for intermediate distance modulus values.  We fix the age and metallicity of the stream to the nearest grid values using the results from Section \ref{sec:cmdfitting}.  For simplicity, we continue to use the \citet{IMFChabrier2001} log-normal IMF and \citet{IMFSalpeter1955} IMF from Section \ref{sec:simCMDs} for $P_{sim}$.  This could affect our ability to trace the stream and the density measurement if there is significant mass segregation, since we expect that low mass stars are stripped first from progenitors, causing a changing mass function along the length of the stream \citep{devil_in_the_tails_Balbinot_2018}.  With age and metallicity fixed, fitting the distance modulus depends primarily on detecting the main sequence and would be independent of the IMF used.   Finally, we create a complete model of all stars by adding the background model to our stream model:

\begin{multline}
    P(\phi_1,\phi_2,c,m|\alpha, I, Pos_{\phi_2}, \sigma, D) = \\
    \alpha \, P_{str} (\phi_1,\phi_2,c,m | Pos_{\phi_2}, \sigma, I, D) + (1-\alpha) \, P_{bg} (\phi_1,\phi_2,c,m)
	\label{eq:spline_model}
\end{multline}

using an additional fitted parameter $\alpha$ that describes the fraction of stream stars among all stars.

\subsection{Model Fitting and Posterior Sampling}
\label{sec:fitting_and_emcee_sampling}

We follow a similar procedure as Section \ref{sec:spatialfitting} to fit our complete model $P(\phi_1,\phi_2,c,m|\alpha, I, Pos_{\phi_2}, \sigma, D)$ to the data.  We calculate the log likelihood of our model for each individual star selected with our matched filter and sum across all stars to compute the total log likelihood ($logL_{total}$).  We use the Nelder-Mead algorithm to find the maximum likelihood model and use a Markov chain Monte Carlo (MCMC) algorithm to sample the posterior and use those results to find the median model with confidence intervals. We perform MCMC sampling using the \textsc{emcee} package \citep{EmceeForeman_Mackey_2013}.  Our number of walkers is four times the number of free parameters of the model being fit to ensure it scales with model complexity and does not need to be adjusted between streams.  We use the first 3000 iterations as  burn-in and discard them.  We then advance the sampler 5000 iterations and use those samples to estimate the model parameters and confidence intervals.  Because our results may be lightly multi-modal, we follow the recommendation in the \textsc{emcee} documentation\footnote{\url{emcee.readthedocs.io}} and use a combination of 80\% differential evolution moves \citep{emcee_DE_moves_Nelson_2013} and 20\% snooker moves using differential evolution \citep{emcee_snooker_DE_moves_ter_braak}.

\subsection{Initial Spline Node Placement}
\label{sec:initial_node_positions}

As our model (Eqn. \ref{eq:spline_model}) is based on cubic splines whose complexity depends on the number of knots/nodes, we need to set the initial number and location of nodes for the four splines.  We base the starting number of spline nodes on the expected complexity of each spline.  Small scale structure is possible and evident in many streams, as seen in \citet{PhoenixDiscoveryBalbinot_2016, SharperViewErkal2017, JhelumMultipleBonaca_2019, closer_look_gd_1_de_Boer_2020, brokenatlasaliqali2020}, resulting in "clumpiness" and significant changes in stellar density across short distances.  To better capture this structure, the density spline is initialized with nodes placed every 1.5 degrees to avoid smoothing over these structures, with at least 7 and up to a maximum of 16 total nodes at the start.  We expect distance modulus, $\phi_2$ position, and width of the stream to vary much more smoothly comparatively.  We start with 3 total nodes for distance modulus, and nodes spaced every 10 (15) degrees for $\phi_2$ ($\sigma$) with at least 5 and up to 12 total nodes for both $\phi_2$ and $\sigma$.

\subsection{Setting the Starting Parameter Values}
\label{sec:setting_starting_parameter_values}

Once we set the number and location of the spline nodes, we need to determine their starting values before we begin adding or removing nodes.  We set these values in two steps.  First, we set the starting values of the nodes based on the results from Section \ref{sec:stage_1_fits}.  We start with a constant density, width, and distance modulus along the length of the stream, with the density values normalised to 1 and the width and distance modulus set to their final values from Section \ref{sec:stage_1_fits}.  Because the coordinate system is aligned with the stream from Section \ref{sec:coordsys}, we initially set $\phi_2 = 0$ for all position nodes.  To reduce computing time, we check by eye if the stream tracks have any noticeable curvature that results in $|\phi_2| > 0.5$.  We then adjust the starting position of the $\phi_2$ nodes so they are within 0.5 degrees of the apparent stream track to avoid starting the search in an area the stream is unlikely to pass through.

From these initial values, we then want to find the best fit values for this number of nodes so we can compare that best fit model to models with different placements of nodes.  To do this, we initially fit our starting model using MCMC following the procedure set out in Section \ref{sec:fitting_and_emcee_sampling} but halving the number of iterations after the burn-in to 2500 to decrease computing time.  Because we use the median value from the MCMC samples, we then use the Nelder-Mead algorithm to find the maximum likelihood model before adding or removing spline nodes.  Only $Pos_{\phi_2}$, I, and $\alpha$ are fit at this time, with the width and distance modulus splines fixed at their initial values.  This focuses the fit on identifying the position and density distribution of the stream first and prevents the model from attempting to fit the background if the starting position and density are too far off their true values, such as with very curved or very clumpy streams.

\subsection{Spline Node Addition/Removal}
\label{sec:add_remove_nodes}

By using cubic splines to model the stream, we can change the complexity of the underlying model to create a more robust program capable of fitting different streams with minimal changes.  Similar to \citet{SharperViewErkal2017}, we use the Akaike Information Criteria \citep[`AIC';][]{AICAkaike1974} to compare models with different sets of spline nodes.  For a given set of nodes, we calculate the corresponding maximum likelihood model using Nelder-Mead.  We can then calculate the AIC value:

\begin{equation}
    AIC = 2k - 2 logL_{total}
	\label{eq:aic_eqn}
\end{equation}

for that model using the number of free parameters (k) and the total log likelihood of the model from Section \ref{sec:fitting_and_emcee_sampling} ($logL_{total}$).  When comparing two models we select the one with the lower AIC value.

We add or remove nodes from a spline one at a time and check multiple positions to determine the best location.  When adding a node, we test placing the new node between pairs of existing nodes at 25\%, 50\%, and 75\%.  When removing nodes, we remove only non-endpoint nodes.  We calculate the AIC value for each test addition or removal, selecting the model with the lowest value.  

Because computational complexity scales rapidly with number of total nodes and the density spline contains the most initial nodes, we first remove nodes from the density spline until the AIC value can no longer be improved.  We then cycle through the distance modulus, density, $Pos_{\phi_2}$, and $\sigma$ splines, attempting to add one node to a spline before proceeding to the next.  We continue to cycle through the four splines until we cannot decrease the AIC value through the addition of any further nodes.  Finally, we repeat the cycle, this time attempting to remove a node from each spline.  With the optimized set of nodes and corresponding maximum likelihood model, we refit the model using MCMC to sample the posterior and determine confidence intervals.
    
\subsection{Model Constraints}
\label{sec:constraints}

When fitting a stream, we set upper and lower bounds for the parameters based on the following reasons:

\begin{itemize}
    \item The distance modulus spline $D(\phi_1)$ is limited to $\pm max(0.3 + 0.01 L, 0.75)$ from the starting distance modulus spline value.  This is based on the extended matched filter range described at the start of Section \ref{sec:stage_2_splines} and ensures the majority of $P_{sim}(c_j, m_j | age, [Fe/H], D)$ will fall within the matched filter.  Limiting the minimum and maximum distance modulus will also effectively limit the possible distance modulus gradient across the stream.
    \item We ensure the stream density $I(\phi_1)$ is positive.
    \item $Pos_{\phi_2}(\phi_1)$ is constrained such that it remains between the minimum $\phi_2$ ($\phi_{2, min}$) and maximum $\phi_2$ ($\phi_{2, max}$) of the region, since we assume the stream is fully contained within our defined region.
    \item We limit $\sigma(\phi_1)$ such that it falls between $0.0143  \leq \sigma(\phi_1) \leq 0.25 (\phi_{2, max}-\phi_{2, min})$ degrees, where $\phi_2$ is also given in degrees.  The lower bound is due to integration resolution limits when normalising $P_{str}(\phi_1,\phi_2 | Pos_{\phi_2}, \sigma, I)$.  Similarly to $Pos_{\phi_2}(\phi_1)$, we set the upper bound based on the assumption that the stream is fully contained within the region.
\end{itemize}

\section{Analysis}
\label{sec:analysis}

In addition to the model of the stream, we also compute several additional position characteristics for each stream that we describe in Section \ref{sec:stream_position_chars}.  We also estimate the total stellar mass and luminosity of the stream according to Section \ref{sec:mass_lum}, using the distance to the stream and fraction of its stellar population we can observe.  Unless otherwise stated, we use the MCMC samples directly to calculate the median value and $16^{th}$ and $84^{th}$ percentiles for these derived values.

\subsection{Stream Position Characteristics}
\label{sec:stream_position_chars}

For comparison to other similar studies, we calculate an orbital pole ($\mathbf{n_{\phi}}$) using the stream track.  This provides a point of reference for each stream to quickly identify a stream's position and orientation.  Using 201 equally spaced points along the stream track ($\mathbf{r_i} = <\phi_{1,i}, Pos_{\phi_2}(\phi_{1,i})>$), we can approximate the total deviation ($\sum_{i=1}^{201}{\arcsin{(\mathbf{n_{\phi}} \cdot \mathbf{r_i})}}$) between the great circle orbit defined by $\mathbf{n_{\phi}}$ and the stream track.  We use the Nelder-Mead algorithm to minimize this deviation and find the best fit for $\mathbf{n_{\phi}}$.  We can then convert $\mathbf{n_{\phi}}$ from $\phi_1$ and $\phi_2$ to RA and Dec using the rotation matrix for that stream.

We also calculate an average width in pc and total length in kpc for each stream.  We first convert the width spline from angular units to physical units using the distance modulus spline.  We then select the longest, unbroken segment of the stream and calculate the average width of that segment.  Because we are calculating the average width along a specific segment, we only use the best fit model and do not calculate confidence intervals with the MCMC samples.  We calculate the length of the stream in kpc by identifying the two endpoints of the stream track and integrating the path in 3d space, using the distance modulus spline to determine the radial distance.

\subsection{Stellar Mass and Luminosity Estimates}
\label{sec:mass_lum}

We estimate the total stellar mass (M) and total luminosity (L) for each stream by essentially reversing the steps from Section \ref{sec:simCMDs}.  We first generate a new simulated stellar population following the same method as Section \ref{sec:simCMDs}, using the best fit age and metallicity values from our initial parameter fits.  Using this population we can calculate the average stellar mass ($M_{ave}$) and luminosity ($L_{ave}$) of the stream stars.  We then divide the stream into 100 segments in $\phi_1$, and calculate the fraction of the simulated population that falls within our matched filter ($f_{match}$) if that population was at the distance modulus of each segment.  We can then calculate M and L for each segment:

\begin{equation}
    M = \frac{M_{ave} N_{tot} I_{seg} \alpha}{f_{match}}
	\label{eq:mass_estimate}
\end{equation}

\begin{equation}
    L = \frac{L_{ave} N_{tot} I_{seg} \alpha}{f_{match}}
	\label{eq:lum_estimate}
\end{equation}

for a segment using the total number of stars after our matched filter ($N_{tot}$), normalized linear density of the segment ($I_{seg}$), and $\alpha$ from our cubic spline model.  Finally, we add the mass and luminosity of each segment to find the total stellar mass and luminosity of the stream.

We determine uncertainties using the uncertainties in $\alpha$ from our sampling of the posterior.  Overall, our method will tend towards underestimating the total stellar mass and luminosity, primarily due to stream visibility, limitations of the mass function we used, and the effect of unresolved binaries.  We often cannot see the full extent of a given stream, whether due to limitations of a survey's footprint, low surface brightness of some parts of the stream, or large gaps isolating parts of the same stream (such as Atlas and Aliqa Uma before they were determined to be the same stream).  This will naturally decrease the calculated mass and luminosity since we would only fit a fraction of the stream.  Additionally, to simplify our model, we assume the mass function along the length of the stream remains constant.  However, low mass stars are typically stripped first, producing a changing mass function along the length of the stream as the stream forms \citep{devil_in_the_tails_Balbinot_2018}.  We will therefore underestimate the number of undetected low mass stars and overestimate $f_{match}$, underestimating the total mass and luminosity.  Our selection of IMF (Section \ref{sec:simCMDs}) will also have a slight effect on our results, but we found for most alternative IMF selections the calculated mass and luminosity differed by less than 0.5\% compared to using just the log-normal IMF or exponential IMF by \citet{IMFChabrier2001}, and less than 2\% compared to the two-part power law IMF by \citet{two_part_power_law_imf_kroupa_2001, two_part_power_law_imf_kroupa_2013}.  We also do not account for unresolved binaries in our model, which produce brighter sequences parallel to the main sequence.  In the case of equal mass binaries, this can increase the apparent magnitude by 0.75 mag \citep{unresolved_binaries_li_2020}, potentially pushing some binaries outside our matched filter and causing us to further underestimate the total stellar mass and luminosity.

\section{Results}
\label{sec:results}

In this section we describe in detail the results of applying the modeling procedure described in previous section to thirteen stellar streams from three surveys.  We fit each stream using only one dataset, following the procedure set out in Section \ref{sec:datasets}.  The resulting node positions and values for all the cubic splines are given in Table \ref{table:node_table}.  We report overall stream characteristics in Table \ref{table:stream_characteristics}, including the calculated stream position characteristics from Section \ref{sec:stream_position_chars} and stellar population information from the initial fits and Section \ref{sec:mass_lum}.  We then discuss the results for each stream in depth in their respective subsections.

\begin{table*}
    \centering
    \begin{tabular}{lllllllll}
        \hline
        \hline
        Stream          & log10(Age/Gyr)     & [Fe/H] & $\alpha_{pole}$ & $\delta_{pole}$ & Total Luminosity       & Stellar Mass           & Length & Width \\
                        &  &     &  (deg)  &  (deg)  & ($L_{\odot}$)          & ($M_{\odot}$)          & (kpc)  & (pc)       \\
        \hline
        \hline
        DES             &         &             &         &          &                        &                        &        &            \\
        \hline
        Aliqa Uma       & 10.00   & -1.1        & 93.3   & 37.5    & $4000_{-370}^{+440}$    & $5240_{-480}^{+580}$    & 5.6    & 210        \\
        ATLAS           & 10.05   & -1.5        & 76.2   & 47.4    & $4950_{-300}^{+270}$    & $6200_{-380}^{+330}$    & 10     & 59         \\
        Chenab          & 10.05   & -1.2        & 254.0  & 13.2    & $15300_{-1200}^{+1200}$ & $20000_{-1500}^{+1600}$ & 24     & 400        \\
        Elqui           & 10.10   & -1.8        & 64.4   & 37.7    & $19800_{-1600}^{+1800}$ & $24100_{-1900}^{+2200}$ & 19     & 550        \\
        Indus           & 10.00   & -1.1        & 24.9   & 21.5    & $13120_{-600}^{+620}$   & $17200_{-790}^{+810}$   & 5.6    & 260        \\
        Jhelum          & 9.95    & -0.9        & 2.6    & 38.0    & $8250_{-460}^{+430}$    & $11010_{-620}^{+570}$   & 5.3    & 240        \\
        Phoenix         & 10.05   & -1.4        & 310.6  & 13.7    & $2750_{-210}^{+210}$    & $3490_{-260}^{+260}$    & 7.4    & 41         \\
        Tucana III      & 10.05   & -1.5        & 353.4  & 30.3    & $3340_{-190}^{+200}$    & $4190_{-230}^{+250}$    & 5.3    & 70         \\
        Turranburra     & 10.10   & -1.6        & 129.9  & 48.4    & $4280_{-360}^{+410}$    & $5340_{-450}^{+510}$    & 8.2    & 250        \\
        Willka Yaku     & 10.00   & -0.8        & 318.1  & 6.0     & $2000_{-230}^{+240}$    & $2760_{-310}^{+330}$    & 8.3   & 83         \\
        \hline
        \hline
        DECaLS          &         &             &         &          &                        &                        &        &            \\
        \hline
        GD 1           & 10.05   & -1.2        & 34.0    & 30.4     & $3450_{-230}^{+250}$    & $4510_{-300}^{+320}$    & 11     & 30         \\
        Pal 5 Leading  & 10.05   & -1.1        & 139.1   & 43.3     & $4490_{-340}^{+340}$    & $5950_{-460}^{+450}$    & 3.9    & 38         \\
        Pal 5 Trailing & 10.05   & -1.1        & 137.8   & 63.6     & $9170_{-320}^{+330}$    & $12150_{-430}^{+430}$   & 9.4    & 45         \\
        Triangulum     & 10.05   & -1.1        & 117.4   & 8.3      & $2700_{-260}^{+250}$    & $3590_{-340}^{+330}$    & 12     & 54         \\
        \hline
        \hline
        Pan-STARRS      &         &             &         &          &                        &                        &        &            \\
        \hline
        Ophiuchus      & 10.00    & -1.9        & 198.9   & 80.6     & $1120_{-88}^{+95}$      & $1330_{-100}^{+110}$    & 1.0    & 9          \\
        \hline
    \end{tabular}
    \caption{Summary of stream characteristics.  Age and metallicity values are best fit values from initial stream fitting and used during spline fit.  Length is measured using the stream track through 3D space from one endpoint of the stream to the other.  Endpoints of the stream are determined by eye.}
    \label{table:stream_characteristics}
\end{table*}

\subsection{Streams in DES}
\label{sec:DES_Streams}

\subsubsection{Aliqa Uma Stream}
\label{sec:Aliqa_Uma}

\begin{figure}
    \centering
    \includegraphics[width=\columnwidth]{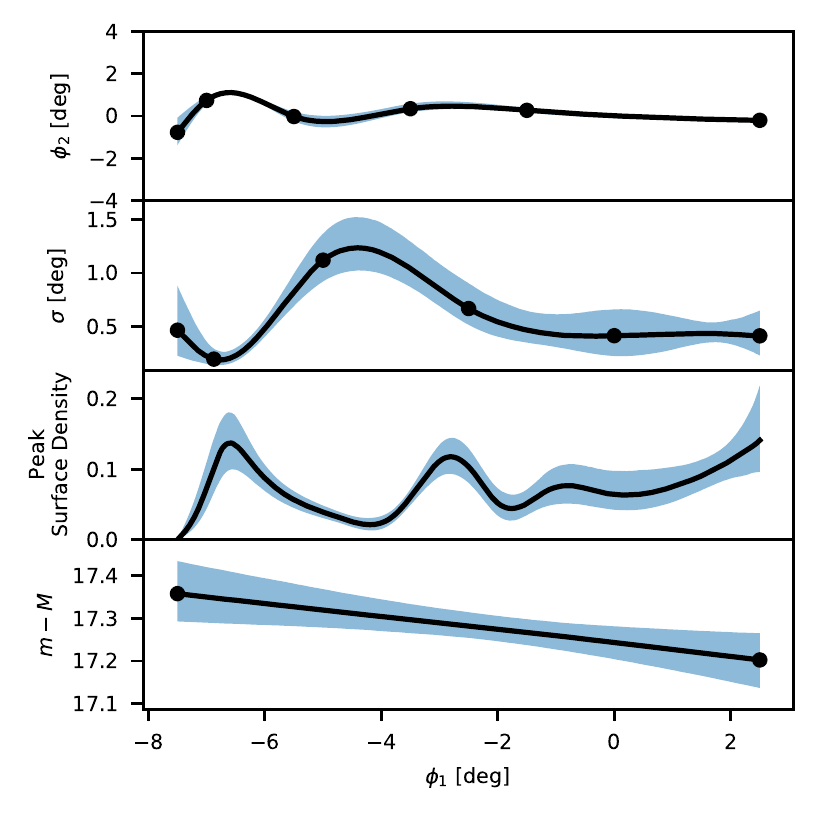}
    \caption{Summary of stream properties for Aliqa Uma from our best fit model, with median (black) and 16th/84th percentiles (shaded blue) curves.  Black points are the final node positions for our model.  The top panel shows stream track ($Pos_{\phi_2}(\phi_1)$ spline). The second panel shows the width of the stream ($\sigma(\phi_1)$ spline).  The third panel shows the peak surface density along the length of the stream.  We derive the peak surface density curve using the linear density spline (I) and the width spline ($\sigma$) from our model.  The bottom panel shows the distance modulus ($D(\phi_1)$ spline).}
    \label{fig:des_aliqa_uma_fit_params}
\end{figure}

\begin{figure}
    \centering
    \includegraphics[width=\columnwidth]{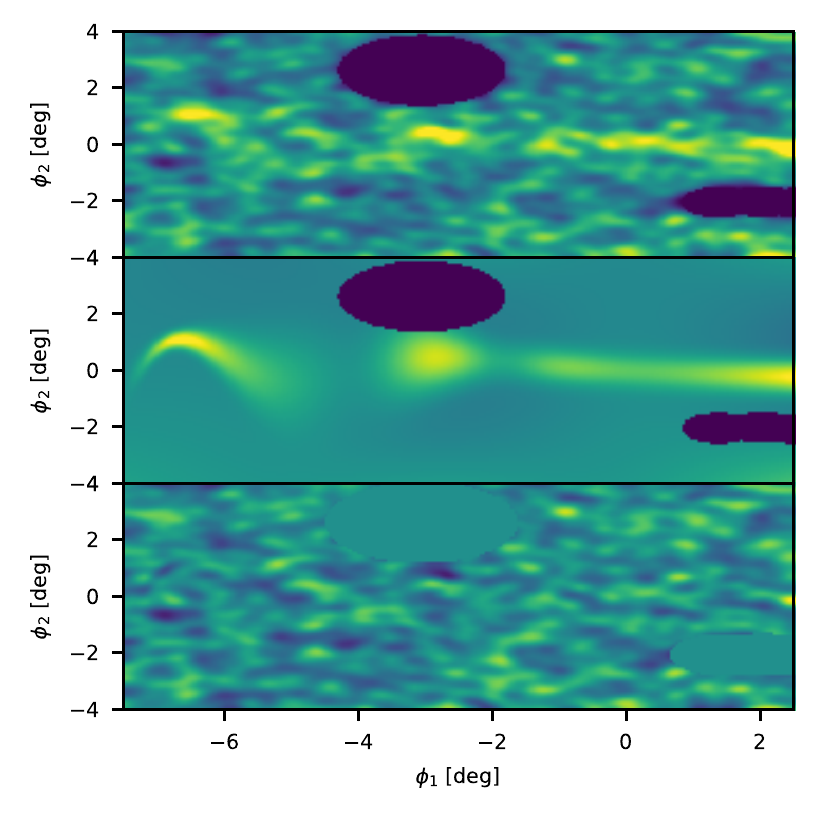}
    \caption{Top: Number density of DES stars near Aliqa Uma with a Gaussian filter with standard deviation of 0.15 degrees applied.  Stars were selected using an matched filter based on a stellar population with $log_{10}(age) = 10.0$, [Fe/H] = -1.1, and distance modulus of 17.14.  Middle:  Best fit model of the stream, normalised to match the number of stars in the top plot.  Bottom: Residual stellar density with colour range centered at 0.}
    \label{fig:des_aliqa_uma_spatial_plots}
\end{figure}

Aliqa Uma is a narrow stream discovered photometrically by \citet{StreamsDESShipp2018} and later found to be an extension of the ATLAS stream by \citet{brokenatlasaliqali2020} using additional spectroscopic observations.  While ATLAS and Aliqa Uma have similar radial velocities and proper motions, there is a distinct "kink" in their stream tracks separating them.  Because of this, we chose to fit ATLAS and Aliqa Uma separately.  The results of our stream analysis of Aliqa Uma are shown in Figures \ref{fig:des_aliqa_uma_fit_params} and \ref{fig:des_aliqa_uma_spatial_plots}.  We focus on a roughly 10 degree segment of the stream, truncating our fit near its overlap with ATLAS to avoid interference.  We also mask the slight overlap with ATLAS near ($\phi_1, \phi_2$) = (2, -2), in addition to Fornax dwarf galaxy near ($\phi_1, \phi_2$) = (-3, 3).  Within this region, we fit a total length of 9.4 degrees ($5.6$ kpc), with an average width of the primary segment from $\phi_1 = -1.5$ to $2.5$ of $0.43$ degrees ($210$ pc).  Figure \ref{fig:des_aliqa_uma_fit_params} shows our best fit model for the stream, where we see a modest distance modulus gradient of $-0.016^{+0.012}_{-0.013}$ mag deg$^{-1}$.  There is a notable over-density at $\phi_1 = -2.8$ followed by a large gap in the stream from $\phi_1 = $ -3.8 to -5.8.  We see a small extension to the stream at $\phi_1 = -6.6$.  Despite the 2 degrees of separation from the main body of the stream, we are confident in this detection due to its alignment with the track of the main segment and its distance modulus.

\begin{figure}
    \centering
    \includegraphics[width=\columnwidth]{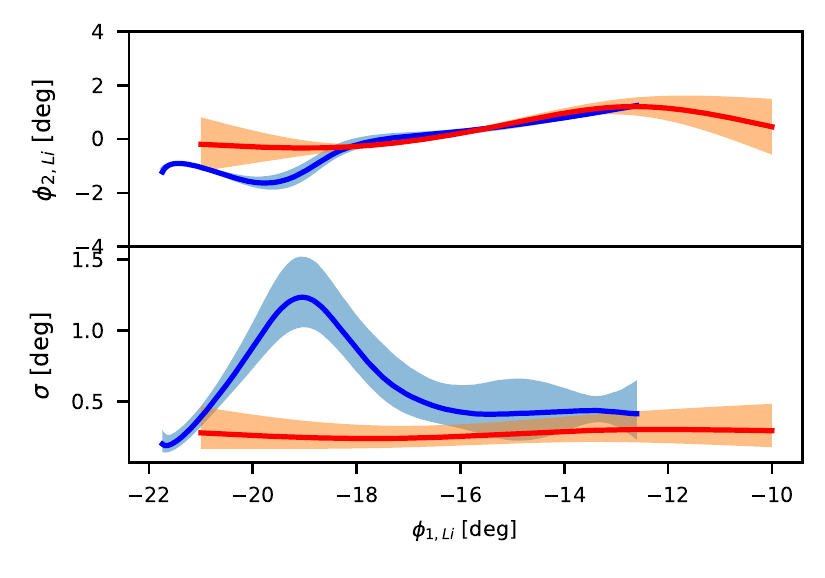}
    \caption{Aliqa Uma stream track (top) and width (bottom) of Aliqa Uma from our model (blue/blue) compared to the spline fit model from \citet{brokenatlasaliqali2020} (red/orange) in their coordinate system. Shaded regions for both plots show the 1-sigma uncertainties of each model.}
    \label{fig:des_aliqa_uma_broken_paper}
\end{figure}

Because of the lack of stream stars in the large gap, we see an artifact of our fitting process at that location as we lose the stream temporarily.  Figure \ref{fig:des_aliqa_uma_broken_paper} highlights this effect when we compare our model to the fit done by \citet{brokenatlasaliqali2020} (using their coordinate system, $\phi_{1,Li}$ and $\phi_{2,Li}$).  The two models are similar, with both stream tracks aligned and a slightly larger width for our model but consistent with their results.  However, we see a clear deviation at $\phi_{1,Li} = -19$ at the location of the gap.  When the stream density drops near zero, the background begins to bias the model and we see the width of the model increase and the stream track appear to curve down.

\subsubsection{ATLAS Stream}
\label{sec:Atlas}

\begin{figure}
    \centering
    \includegraphics[width=\columnwidth]{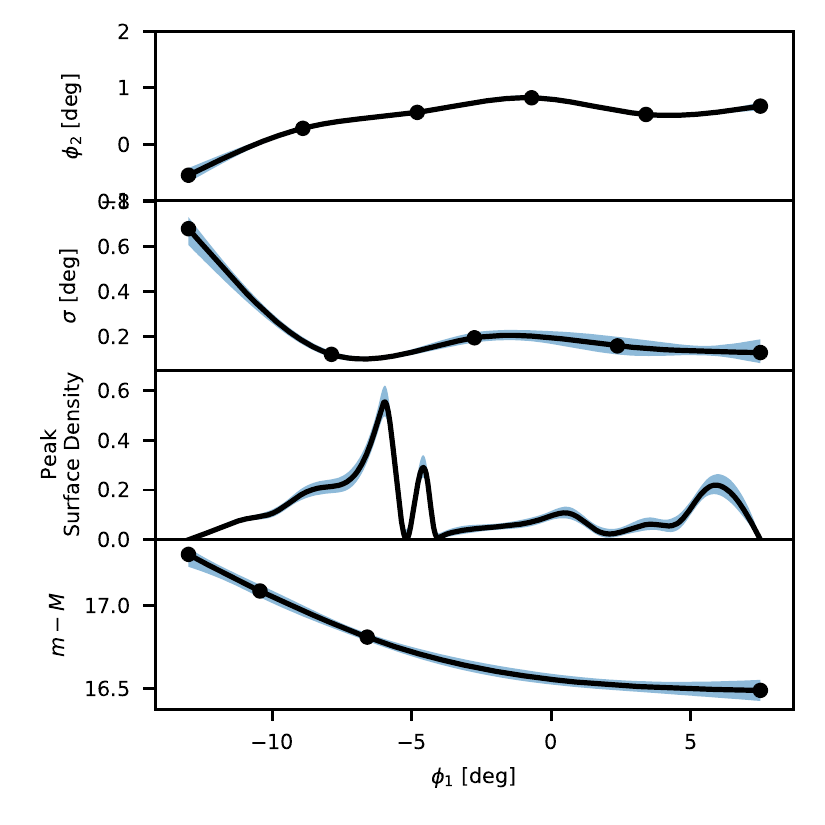}
    \caption{Summary of stream properties for ATLAS from our best fit model, with median (black) and 16th/84th percentiles (shaded blue) curves.  Black points are the final node positions for our model.} 
    \label{fig:des_atlas_fit_params}
\end{figure}

\begin{figure}
    \centering
    \includegraphics[width=\columnwidth]{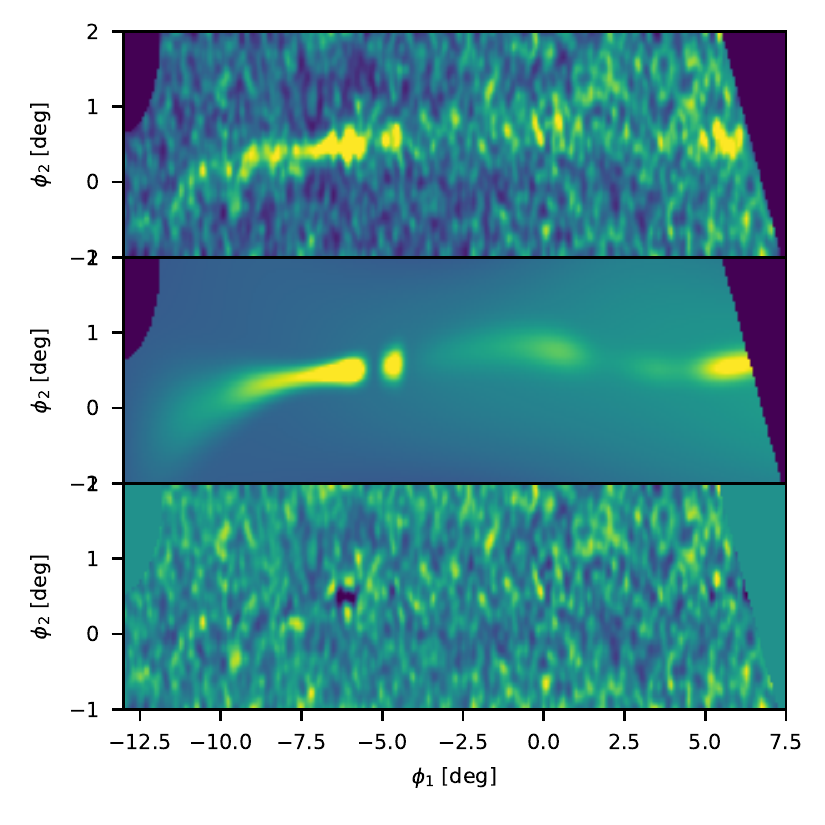}
    \caption{Top: Number density of DES stars near ATLAS with a Gaussian filter with standard deviation of 0.075 degrees applied.  Stars were selected using an matched filter based on a stellar population with $log_{10}(age) = 10.05$, [Fe/H] = -1.5, and distance modulus of 16.76.  Middle:  Best fit model of the stream, normalised to match the number of stars in the top plot.  Bottom: Residual stellar density with colour range centered at 0.}
    \label{fig:des_atlas_spatial_plots}
\end{figure}

The ATLAS stream is a thin stellar stream first discovered in the ATLAS survey by \citet{Atlas_discovery_Koposov_2014}.  Follow-up measurements were done by \citet{StreamsDESShipp2018} using DES, and \citet{brokenatlasaliqali2020} later linked ATLAS to Aliqa Uma using spectroscopy from the S5 survey.  The results of our stream analysis are shown in Figures \ref{fig:des_atlas_fit_params} and \ref{fig:des_atlas_spatial_plots}.  We take a similar approach with ATLAS as we did with Aliqa Uma; this time we mask Aliqa Uma in the upper left of Figure \ref{fig:des_atlas_spatial_plots} to avoid interference.  On the upper right side we encounter the edge of the DES dataset.  We fit a total length of 18 degrees (10 kpc) with an average width of 0.16 degrees (59 pc), although the stream continues past the edge of the dataset.  There is a notable distance modulus gradient along the length of ATLAS that averages $-0.036^{+0.003}_{-0.003}$ mag deg$^{-1}$ ($-0.38^{+0.03}_{-0.03}$ kpc deg$^{-1}$).  This value is in the middle of the range of gradients seen by \citet{brokenatlasaliqali2020}.  Their  measurements using spectroscopically confirmed BHB stars showed the distance gradient change slightly along the length of the stream, with a best fit distance gradient of $-0.20$ kpc deg$^{-1}$ at $\phi_1 = 7$ and $-0.43$ kpc deg$^{-1}$ at $\phi_1 = -10$ .  We note two large over-densities at $\phi_1 =$ -6.0 and -4.6 with a pronounced gap separating them that contains almost no stream stars (Figure \ref{fig:des_atlas_fit_params}).  We detect part of a third over-density around $\phi_1 = 7$ that appears to continue past the edge of the dataset.  The segment of the stream from $\phi_1 = -4$ to $\phi_1 = 5$ has a lower stellar density, with a slight increase in density at $\phi_1 = 0$ and slight broadening of the stream over that area.

\begin{figure}
    \centering
    \includegraphics[width=\columnwidth]{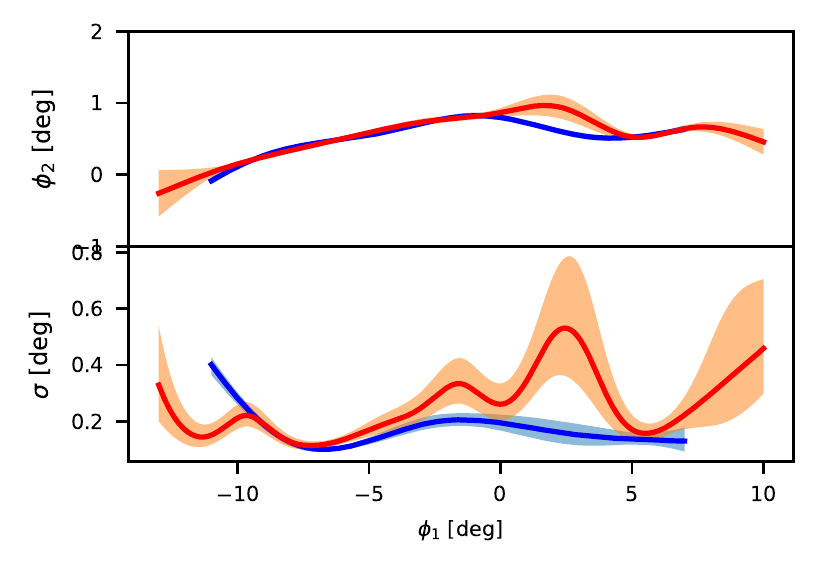}
    \caption{Atlas stream track (top) and width (bottom) of our model (blue/blue) compared to the spline fit model from \citet{brokenatlasaliqali2020} (red/orange) in their coordinate system. Shaded regions show the 1-sigma uncertainties for each model.}
    \label{fig:des_atlas_broken_paper}
\end{figure}

Our density, width, and stream track are in general agreement with the model from \citet{brokenatlasaliqali2020}.  There are three main differences between our models, primarily arising due to differences in how they are designed.  We focused primarily on detailed linear density fitting, with a sufficiently robust model to fit the track, width, and distance modulus with minimal overfitting.  As a result, we were able to resolve the single over-density near $\phi_1 = -5$ in \citet{brokenatlasaliqali2020} into the two previously mentioned over-densities at $\phi_1 =$ -6.0 and -4.6.  However, as shown in Figure \ref{fig:des_atlas_broken_paper}, we over-smoothed the width of the stream and do not detect the wiggle broadening that \citet{brokenatlasaliqali2020} finds at $\phi_1 = 2.5$.  Our model also begins to extrapolate the width increase seen at $\phi_1 = -10$ as we reach the end of the stream.  Otherwise our stream and width track agrees with the measurements by \citet{brokenatlasaliqali2020}.  We also observe similar over-densities as they do at $\phi_1 = 0.5$ and $\phi_1 = 7$.

\subsubsection{Chenab Stream}
\begin{figure}
    \centering
    \includegraphics[width=\columnwidth]{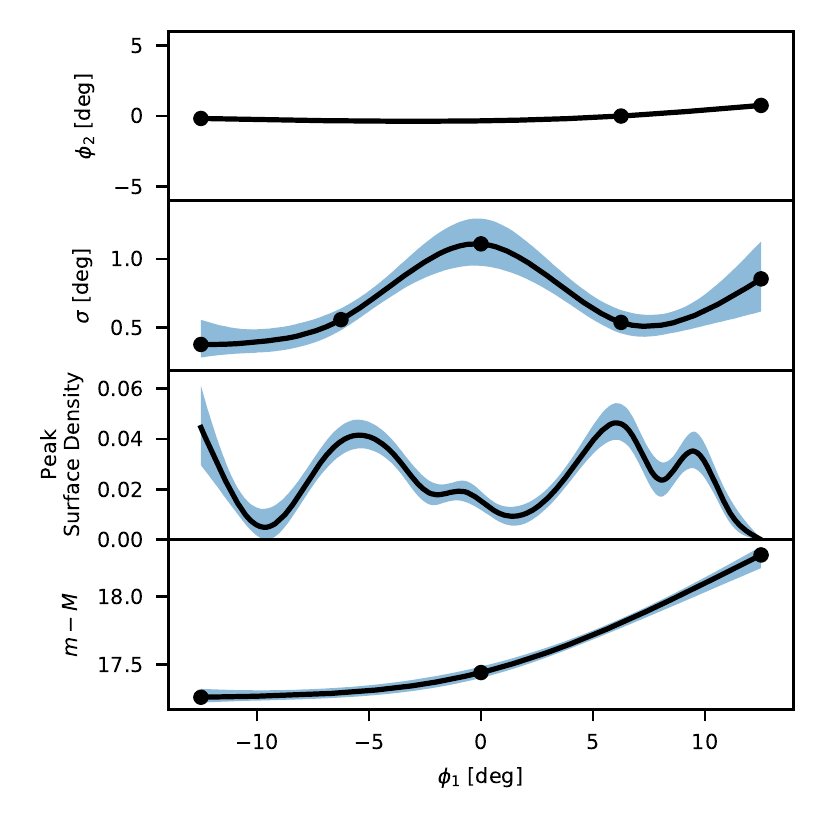}
    \caption{Summary of stream properties for Chenab from our best fit model, with median (black) and 16th/84th percentiles (shaded blue) curves.  Black points are the final node positions for our model.}
    \label{fig:des_chenab_fit_params}
\end{figure}

\begin{figure}
    \centering
    \includegraphics[width=\columnwidth]{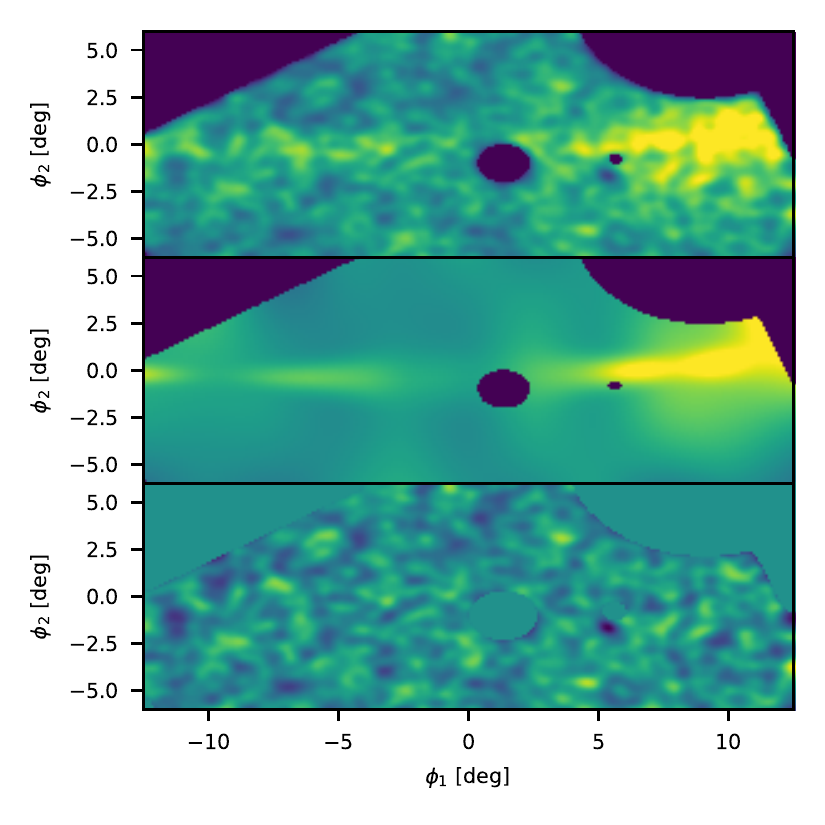}
    \caption{Top: Number density of DES stars near Chenab with a Gaussian filter with standard deviation of 0.25 degrees applied.  Stars were selected using an matched filter based on a stellar population with $log_{10}(age) = 10.05$, [Fe/H] = -1.2, and distance modulus of 17.8.  Middle:  Best fit model of the stream, normalised to match the number of stars in the top plot.  Bottom: Residual stellar density with colour range centered at 0.}
    \label{fig:des_chenab_spatial_plots}
\end{figure}

Chenab is a fainter stellar stream and part of the group of streams first discovered by \citet{StreamsDESShipp2018}.  Since its discovery, it has been identified as an extension to the Orphan stream by \citet{ChenabOrphanKoposov_2019}.  The results of our stream analysis are shown in Figures \ref{fig:des_chenab_fit_params} and \ref{fig:des_chenab_spatial_plots}.  When analysing the stream we masked the nearby Grus II dwarf galaxy at $(\phi_1, \phi_2) = (6, -1)$ \citep{Eight_Ultrafaint_Drlica-Wagner2015}.  Additionally, limited survey coverage near the stream is evident in three areas: near the center of the stream at $(\phi_1, \phi_2) = (1.5, -1)$ and in the upper left and upper right of Figure \ref{fig:des_chenab_spatial_plots}.  The stream extends past the edge of the dataset on both sides, with the visible portion of the stream spanning 25 degrees (24 kpc) with an average width of 0.7 degrees (400 pc).  For $\phi_1 < 0$ there is minimal change in distance modulus while for $\phi_1 > 0$ we see a significant gradient of $0.069^{+0.010}_{-0.010}$ mag deg$^{-1}$. The gap in DES data at $\phi_1 = 1.5$ limits our analysis of the density profile of the stream as the stream track passes directly over it.  However, away from the gap in the footprint, we see comparable peak surface densities in Figure \ref{fig:des_chenab_fit_params} for $\phi_1 < -2$ and $\phi_1 > 4$ suggesting a fairly uniform density along the length of the stream with some possible under-densities at $\phi_1 = -9$ and $\phi_1 = 8$.

\subsubsection{Elqui Stream}

\begin{figure}
    \centering
    \includegraphics[width=\columnwidth]{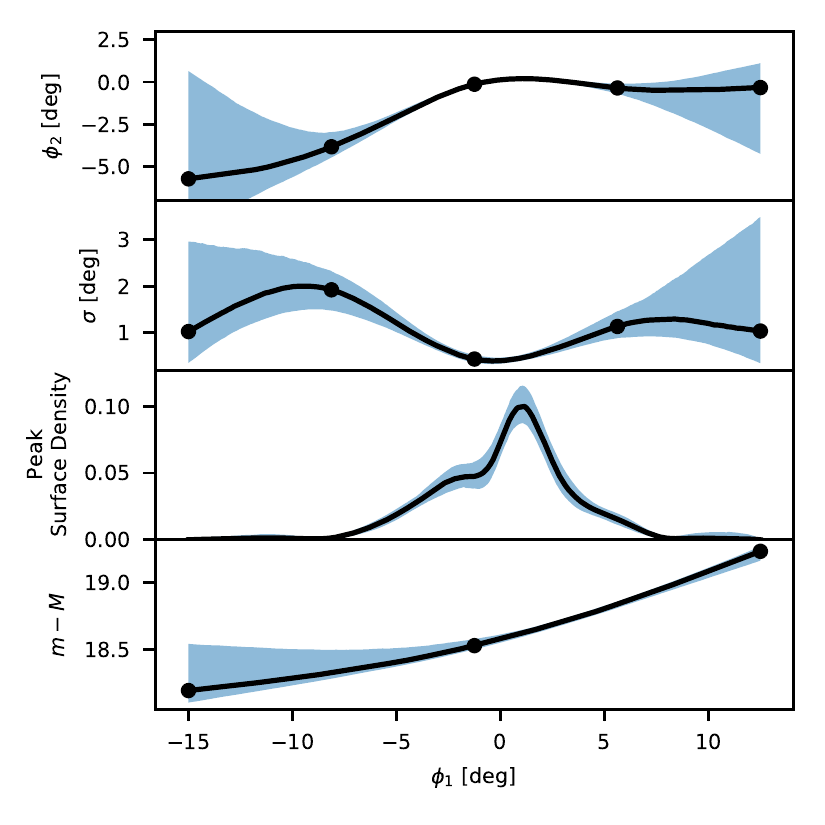}
    \caption{Summary of stream properties for Elqui from our best fit model, with median (black) and 16th/84th percentiles (shaded blue) curves.  Black points are the final node positions for our model.}
    \label{fig:des_elqui_fit_params}
\end{figure}

\begin{figure}
    \centering
    \includegraphics[width=\columnwidth]{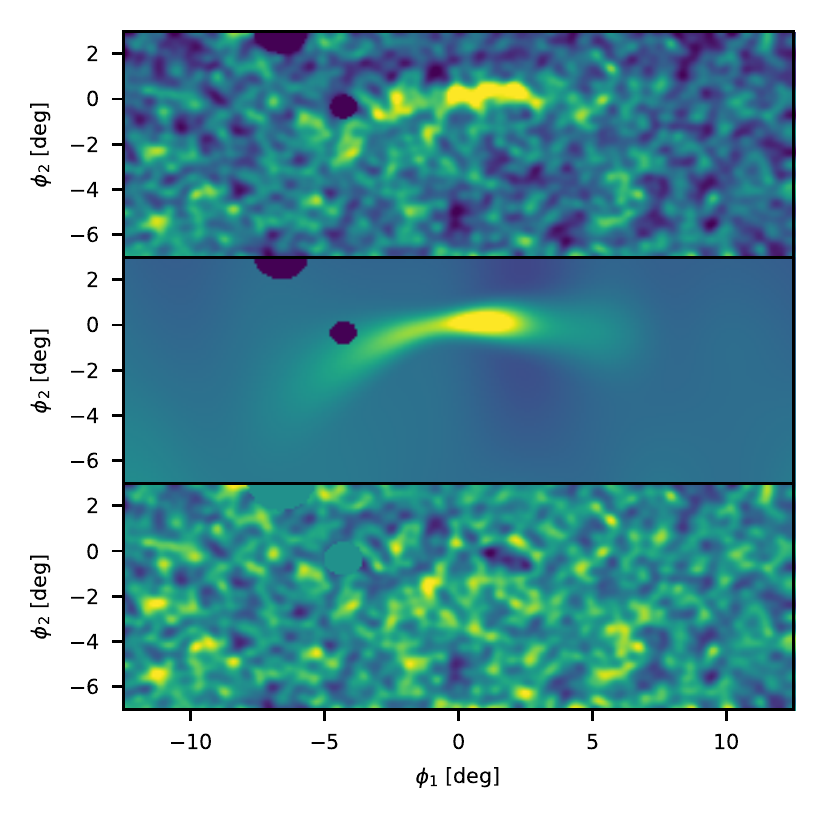}
    \caption{Top: Number density of DES stars near Elqui with a Gaussian filter with standard deviation of 0.2 degrees applied.  Stars were selected using an matched filter based on a stellar population with $log_{10}(age) = 10.10$, [Fe/H] = -1.8, and distance modulus of 18.68.  Middle:  Best fit model of the stream, normalised to match the number of stars in the top plot.  Bottom: Residual stellar density with colour range centered at 0.}
    \label{fig:des_elqui_spatial_plots}
\end{figure}

The most distant stream discovered by \citet{StreamsDESShipp2018} at 50.1 kpc, Elqui is probably near the edge of what can be detected with DES.  The results of our stream analysis are shown in Figures \ref{fig:des_elqui_fit_params} and \ref{fig:des_elqui_spatial_plots}.  We mask NGC 300 and Sculptor at $(\phi_1, \phi_2) = (-4, -0.5)$ and $(-7, 3)$ respectively.  We found the stream to be 14 degrees (19 kpc) long with an average width of 0.6 degrees (550 pc), and see the same curvature in the stream track past $\phi_1 < -4$ as \cite{StreamsDESShipp2018} (Figure \ref{fig:des_elqui_fit_params}).  We also see a significant distance modulus gradient of $0.038^{+0.004}_{-0.012}$ mag deg$^{-1}$.  The large peak in density and narrow width at $\phi_1 = 1$ compared to the rest of the stream suggests a possible location of the progenitor.  We see similar behavior with the Tucana III and Palomar 5 streams, where the stream begins to narrow near the progenitor.  However, we do not see a discontinuity in the stream track at $\phi_1 = 1$ which we might expect at a progenitor, such as the one that occurs for Pal 5.

\subsubsection{Indus Stream}

\begin{figure}
    \centering
    \includegraphics[width=\columnwidth]{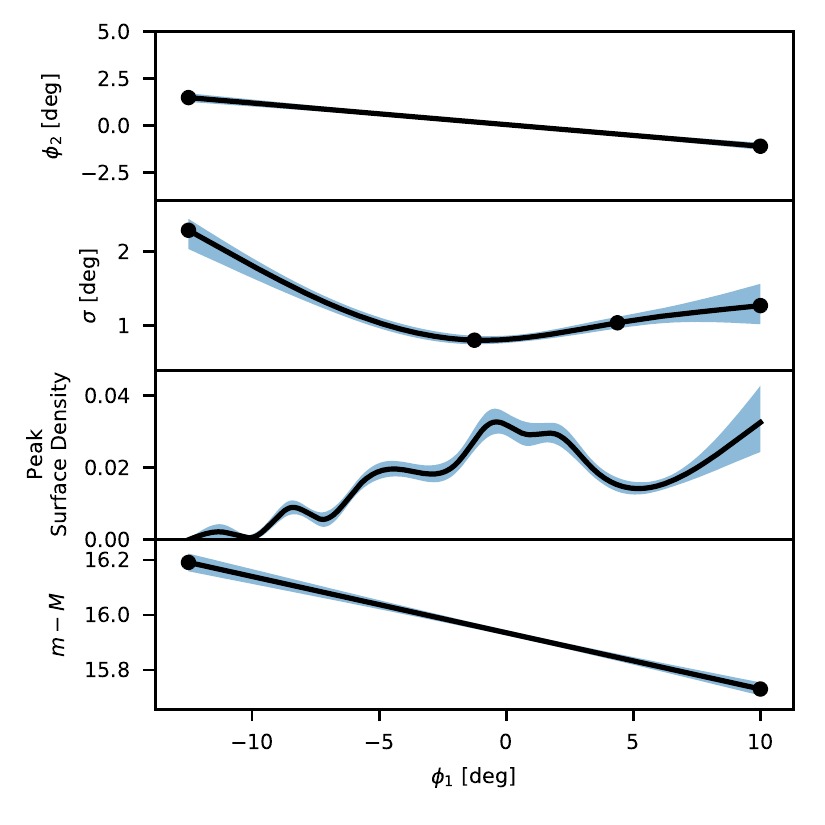}
    \caption{Summary of stream properties for Indus from our best fit model, with median (black) and 16th/84th percentiles (shaded blue) curves.  Black points are the final node positions for our model.}
    \label{fig:des_indus_fit_params}
\end{figure}

\begin{figure}
    \centering
    \includegraphics[width=\columnwidth]{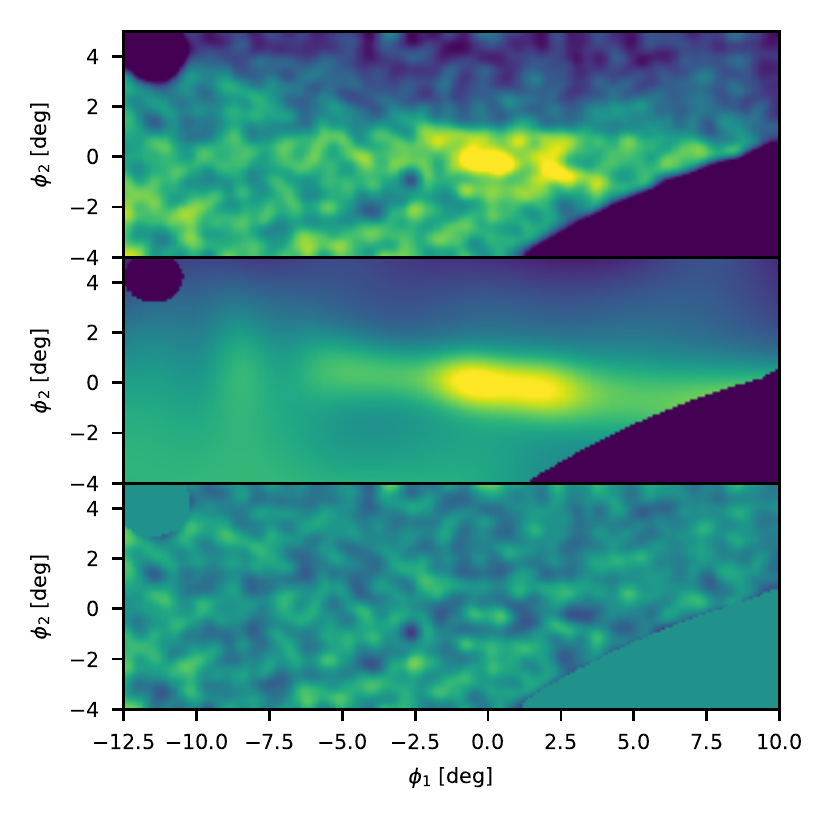}
    \caption{Top: Number density of DES stars near Indus with a Gaussian filter with standard deviation of 0.25 degrees applied.  Stars were selected using an matched filter based on a stellar population with $log_{10}(age) = 10.0$, [Fe/H] = -1.1, and distance modulus of 15.9  Middle:  Best fit model of the stream, normalised to match the number of stars in the top plot.  Bottom: Residual stellar density with colour range centered at 0.}
    \label{fig:des_indus_spatial_plots}
\end{figure}

Indus is another stellar stream first identified by \citet{StreamsDESShipp2018}, located near Tucana III dwarf galaxy and Jhelum stream.  The results of our stream analysis are shown in Figures \ref{fig:des_indus_fit_params} and \ref{fig:des_indus_spatial_plots}.  At the bottom right of Figure \ref{fig:des_indus_spatial_plots} is the edge of the DES footprint and contains no data, with an additional hole in the DES data located in the top left at $(\phi_1, \phi_2) = (-11, 4)$. Because of the edge of the dataset, we are only able to see a segment of the stream 18.5 degrees (5.6 kpc) long with an average width of 1.0 degrees (260 pc).  We see a linear stream track with a slight distance modulus gradient, averaging $-0.020^{+0.002}_{-0.002}$ mag deg$^{-1}$ along its length (Figure \ref{fig:des_indus_fit_params}).  While limited by the edge of the footprint, we are able to still identify some notable features in the stream.  There is a primary group of stars from $\phi_1 = -2$ to 3, with lower density tails on either side stretching from $\phi_1 = -6$ to -2 and from $\phi_1 = 3$ to the edge of the footprint.  These lower density features have roughly half the peak surface density as the central clump, with sharp changes in density between them (Figure \ref{fig:des_indus_spatial_plots}).

\subsubsection{Jhelum Stream}

\begin{figure}
    \centering
    \includegraphics[width=\columnwidth]{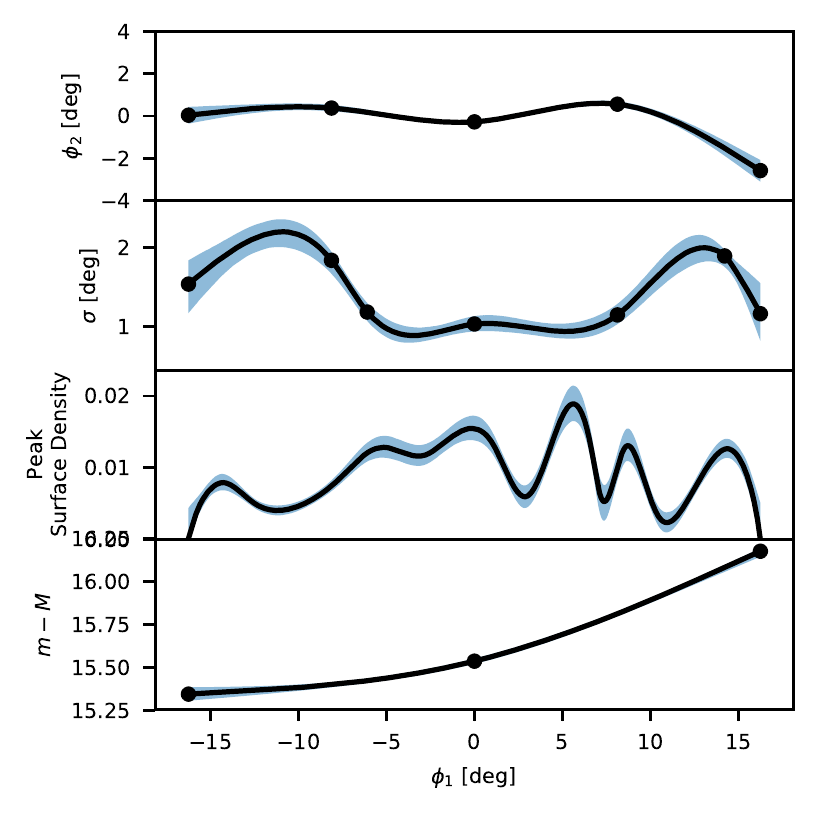}
    \caption{Summary of stream properties for Jhelum from our best fit model, with median (black) and 16th/84th percentiles (shaded blue) curves.  Black points are the final node positions for our model.}
    \label{fig:des_jhelum_fit_params}
\end{figure}

\begin{figure}
    \centering
    \includegraphics[width=\columnwidth]{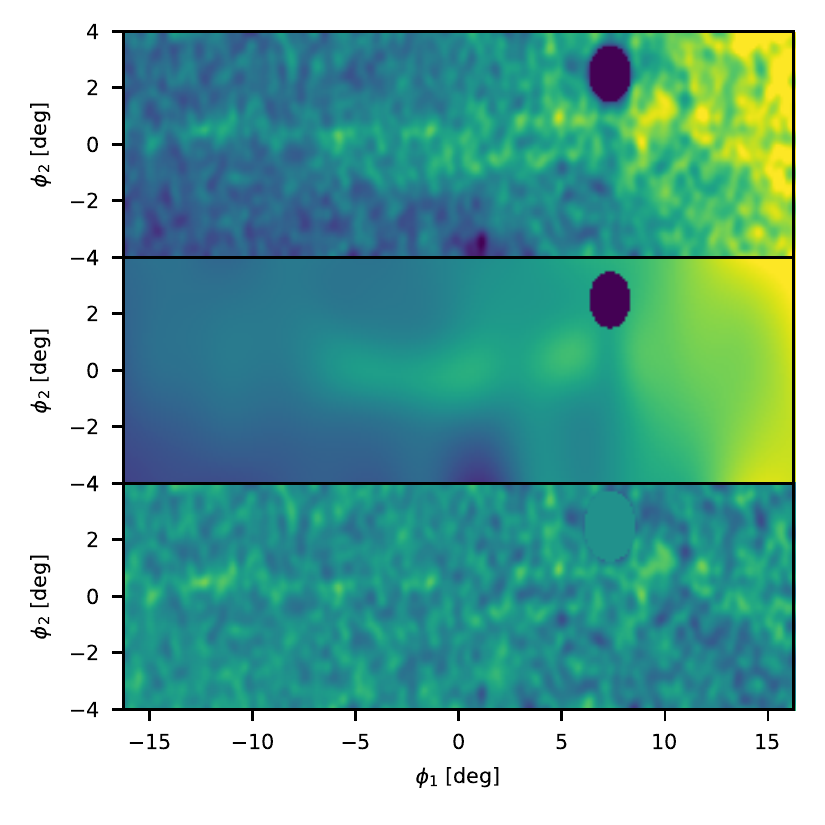}
    \caption{Top: Number density of DES stars near Jhelum with a Gaussian filter with standard deviation of 0.2 degrees applied.  Stars were selected using an matched filter based on a stellar population with $log_{10}(age) = 9.95$, [Fe/H] = -0.9, and distance modulus of 15.5.  Middle:  Best fit model of the stream, normalised to match the number of stars in the top plot.  Bottom: Residual stellar density with colour range centered at 0.}
    \label{fig:des_jhelum_spatial_plots}
\end{figure}

A rather complex stream, \citet{StreamsDESShipp2018} first discovered Jhelum in the western half of DES, just North of Tucana III dwarf and Indus stream. Further analysis by \citet{JhelumMultipleBonaca_2019} identified narrow and broad components of the stream along the majority of its length.  As a result, we struggle to fit Jhelum with our model since we cannot divide the stream into multiple components.  We focus primarily on the wide component of the stream, constraining the width of the stream using a Gaussian prior on log($\sigma [deg]$) with a mean at -0.0725 (0.93 degrees) and standard deviation of 0.2 and slightly constraining the position of the stream using a Gaussian prior with a mean at 0 and standard deviation of 1.0 degrees.  The results of our analysis are shown in Figures \ref{fig:des_jhelum_fit_params} and \ref{fig:des_jhelum_spatial_plots}.  Despite these limitations, we can still see some structure within the stream.  We see the broad component of the stream stretches from $\phi_1 = -6$ to $\phi_1 = 8$, with slight changes in density (Figure \ref{fig:des_jhelum_fit_params}).  Additionally, there is an average distance modulus gradient of $0.014^{+0.003}_{-0.003}$ mag deg$^{-1}$ along its length.  We were also able to indirectly confirm the thin component of Jhelum, visible in the residual plot of Figure \ref{fig:des_jhelum_spatial_plots} stretching from $\phi_1 =$ -15 to 10 at $\phi_2 = 0.5$.  We see the same diffuse background over-density as \citet{StreamsDESShipp2018} past $\phi_1 > 10$, limiting our ability to fit the stream there without additional data to separate out background stars, such as proper motions used by \citet{JhelumMultipleBonaca_2019}.

\subsubsection{Phoenix Stream}

\begin{figure}
    \centering
    \includegraphics[width=\columnwidth]{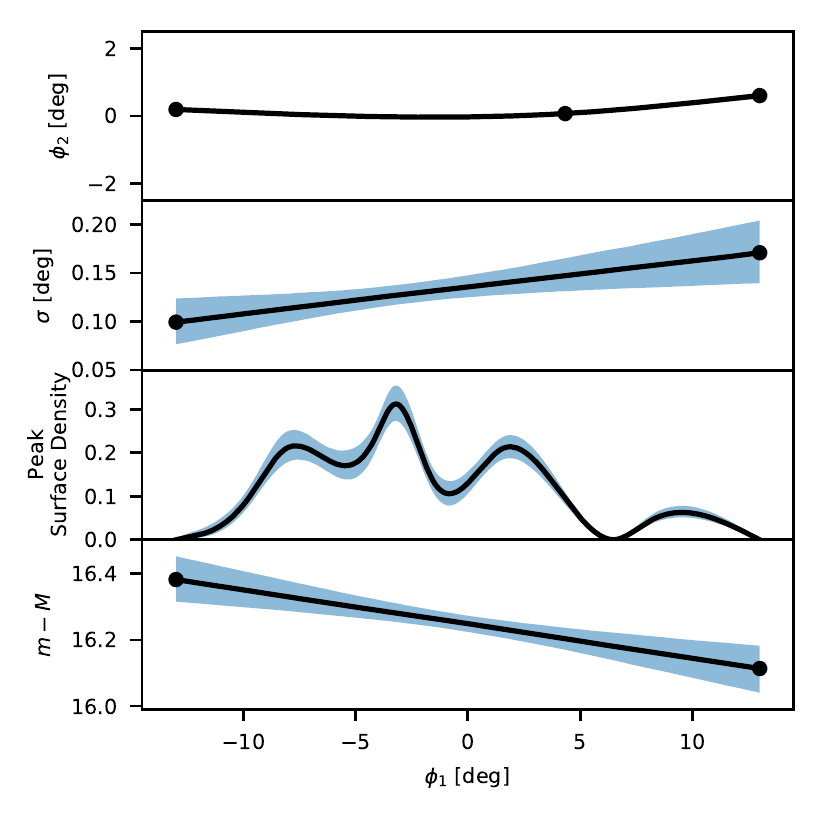}
    \caption{Summary of stream properties for Phoenix stream from our best fit model, with median (black) and 16th/84th percentiles (shaded blue) curves.  Black points are the final node positions for our model.}
    \label{fig:des_phoenix_fit_params}
\end{figure}

\begin{figure}
    \centering
    \includegraphics[width=\columnwidth]{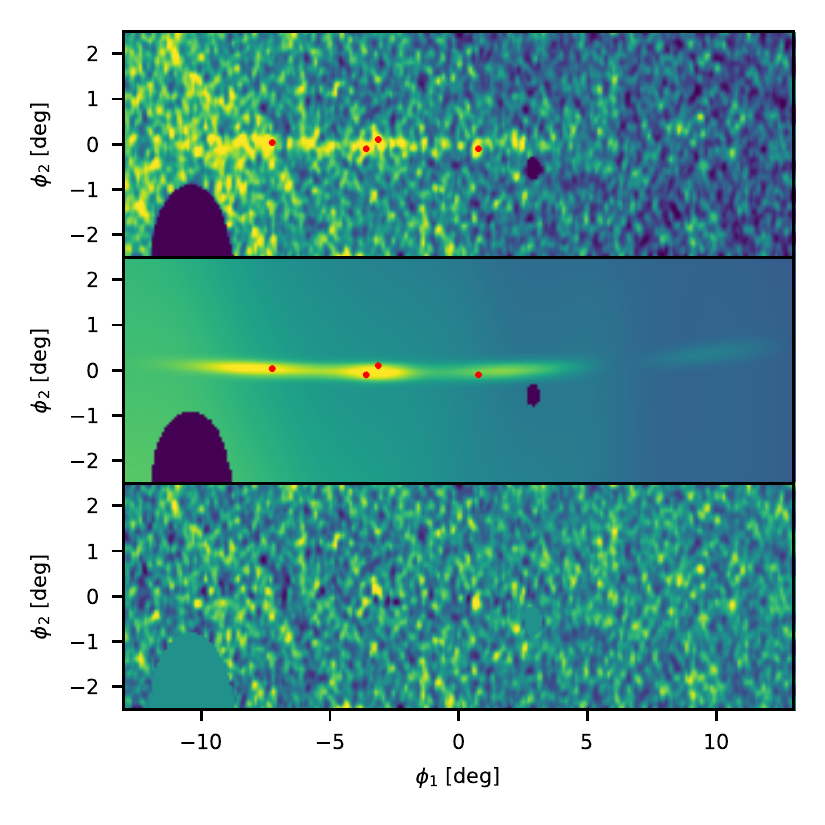}
    \caption{Top: Number density of DES stars near Phoenix with a Gaussian filter with standard deviation of 0.075 degrees applied.  Stars were selected using an matched filter based on a stellar population with $log_{10}(age) = 10.05$, [Fe/H] = -1.4, and distance modulus of 16.38.  Red points are the over-densities noted in \citet{PhoenixDiscoveryBalbinot_2016} (S, C1, C2, and N from left to right).  Middle:  Best fit model of the stream, normalised to match the number of stars in the top plot.  Bottom: Residual stellar density with colour range centered at 0.}
    \label{fig:des_phoenix_spatial_plots}
\end{figure}

\begin{figure}
    \centering
    \includegraphics[width=\columnwidth]{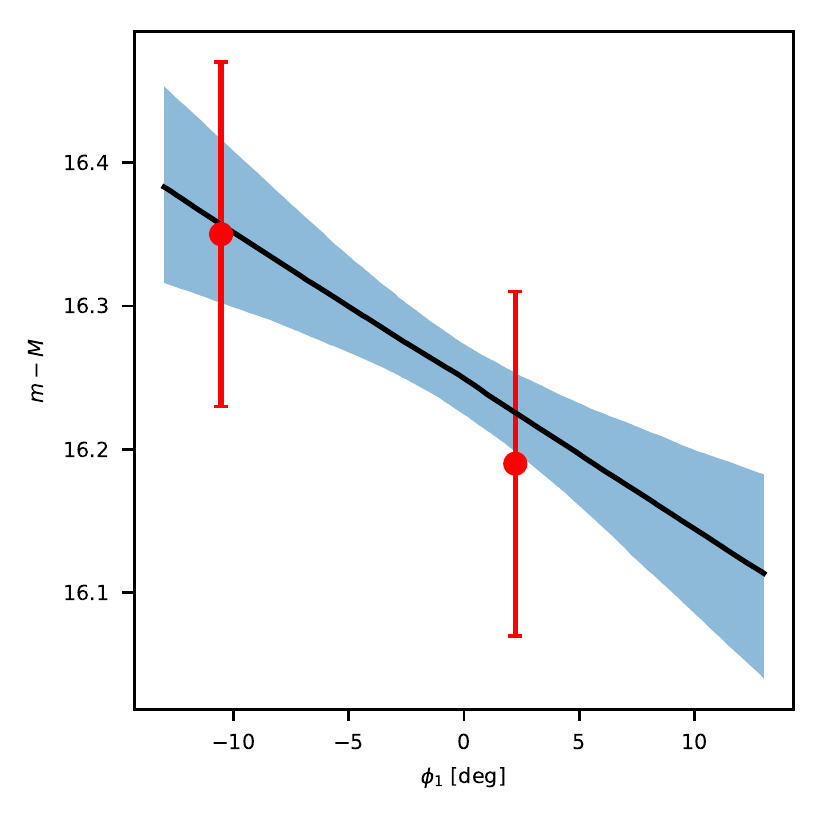}
    \caption{Distance modulus for Phoenix stream, with our model (black/blue) compared to measurements by \citet{PhoenixDiscoveryBalbinot_2016} (red)}
    \label{fig:des_phoenix_dist_mod_compare_plot}
\end{figure}

First discovered and analyzed by \citet{PhoenixDiscoveryBalbinot_2016} and followed up by \citet{StreamsDESShipp2018}, the Phoenix stream is located next to the Phoenix dwarf galaxy \citep{PhoenixDwarfCanterna1977} (masked at $(\phi_1, \phi_2) = (3, -0.5)$).  In addition to the Phoenix dwarf, there is an empty region at $(\phi_1, \phi_2) = (-10, -2)$ due to a gap in DES data.  The results of our stream analysis are shown in Figures \ref{fig:des_phoenix_fit_params} and \ref{fig:des_phoenix_spatial_plots}.  We see a fairly constant width of 0.13 degrees (41 pc) along its 23 degree (7.4 kpc) length.  We calculate a distance modulus gradient of $-0.010^{+0.005}_{-0.005}$ mag deg$^{-1}$, comparable to the $-0.02^{+0.02}_{-0.02}$ mag deg$^{-1}$ gradient noted by \citet{PhoenixDiscoveryBalbinot_2016} (Figure \ref{fig:des_phoenix_dist_mod_compare_plot}).  \citet{PhoenixDiscoveryBalbinot_2016} also identifies two central over-densities clumped together, C1 and C2 in their paper, and a North and a South over-density, each located approximately 4 degrees from the two central over-densities (Figure \ref{fig:des_phoenix_spatial_plots}).  We see the same three clumps but smooth slightly over the entirety of the stream and cannot individually resolve C1 and C2 (Figure \ref{fig:des_phoenix_fit_params}).  We see a potential extension to the stream roughly 5 degrees further north than the previously known edge of the stream, near $\phi_1 = 10$.  This group of stars deviates slightly from the main stream track  by roughly 0.5 degrees, but maintains the same distance modulus gradient.  However, the close proximity of the Palca and Turbio streams in that region, in addition to the low surface density and separation from the rest of the Phoenix stream, makes it difficult to confidently label it an extension to the stream.  A follow up study that includes proper motions or uses spectroscopy to identify member stars would be necessary to determine if this additional group of stars at $\phi_1 = 10$ is part of the stream.

\subsubsection{Tucana III Stream}

\begin{figure}
    \centering
    \includegraphics[width=\columnwidth]{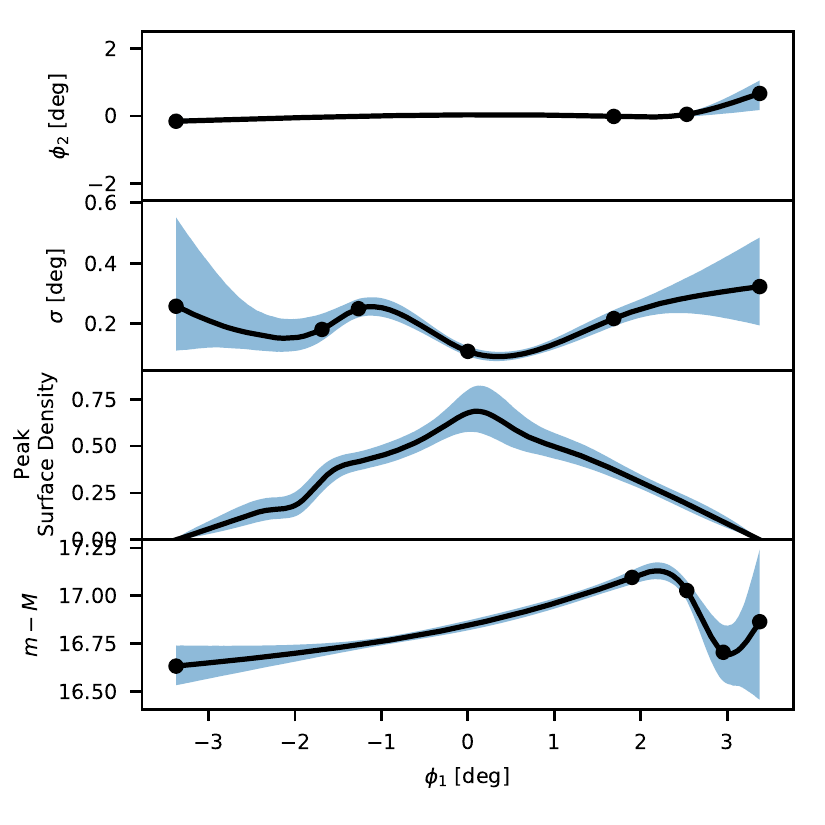}
    \caption{Summary of stream properties for Tucana III from our best fit model, with median (black) and 16th/84th percentiles (shaded blue) curves.  Black points are the final node positions for our model.}
    \label{fig:des_tucana_iii_fit_params}
\end{figure}

\begin{figure}
    \centering
    \includegraphics[width=\columnwidth]{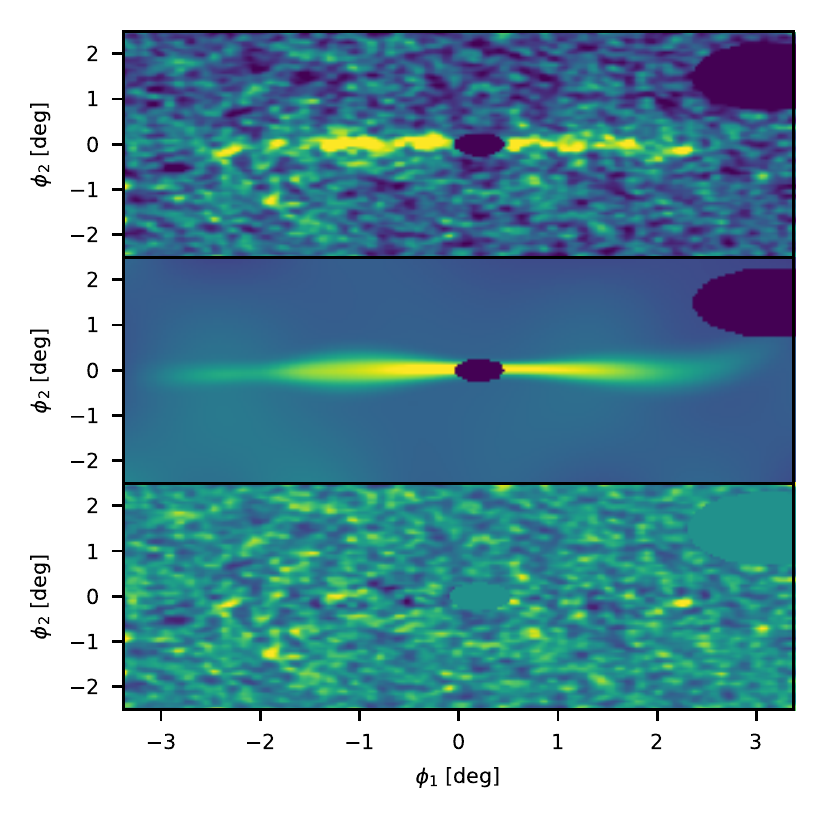}
    \caption{Top: Number density of DES stars near Tucana III with a Gaussian filter with standard deviation of 0.05 degrees applied.  Stars were selected using an matched filter based on a stellar population with $log_{10}(age) = 10.05$, [Fe/H] = -1.5, and distance modulus of 16.84.  Middle:  Best fit model of the stream, normalised to match the number of stars in the top plot.  Bottom: Residual stellar density with colour range centered at 0.}
    \label{fig:des_tucana_iii_spatial_plots}
\end{figure}

\begin{figure}
    \centering
    \includegraphics[width=\columnwidth]{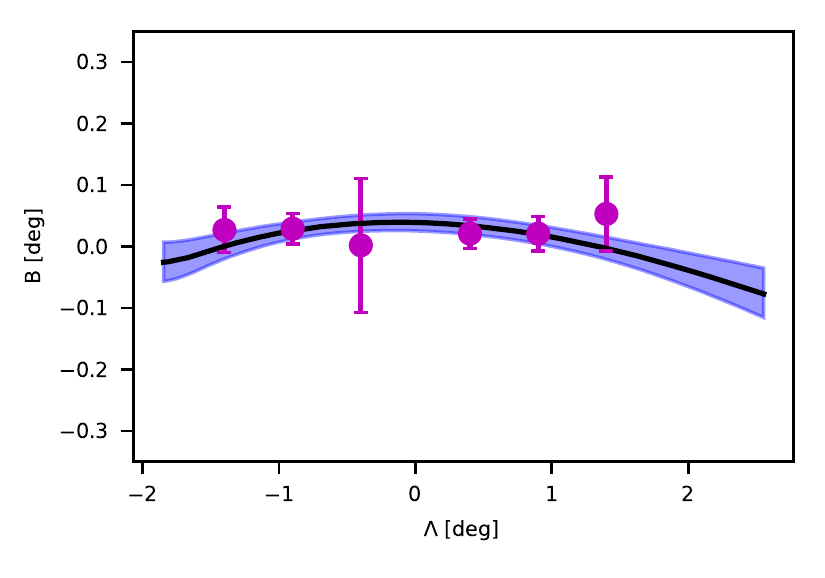}
    \caption{Tucana III stream track in rotated, stream aligned coordinates from \citet{TucanaIIIErkal_2018}.  Black/blue: Our model of the stream track with 1-sigma uncertainties in blue. Magenta: Stream track from \citet{TucanaIIIErkal_2018}.}
    \label{fig:des_tucana_iii_spatial_compare_plot_lb}
\end{figure}

Tucana III is an ultra-faint dwarf galaxy initially discovered by \citet{Eight_Ultrafaint_Drlica-Wagner2015}, who also noted extended tails extending approx. 2 degrees from either side.  \citet{StreamsDESShipp2018} followed up on their observations, confirming the tails and also finding a distance modulus gradient of $0.16^{+0.06}_{-0.06}$ mag deg$^{-1}$.  For our fit, we mask Tucana III itself to focus our model on the tidal tails, and also mask part of a diffuse stellar over-density (top right of the panels in Figure \ref{fig:des_tucana_iii_spatial_plots}) noted in \citet{StreamsDESShipp2018}. The results of our fit are shown in Figures \ref{fig:des_tucana_iii_fit_params} and \ref{fig:des_tucana_iii_spatial_plots}.  We calculate the total length of the tidal tails to be 4.4 degrees (5.3 kpc) and an average width of 0.17 degrees (70 pc).  Our stream track is consistent with the measured stream track from \citet{TucanaIIIErkal_2018}, with minimal curvature along the length of the stream (Figure \ref{fig:des_tucana_iii_spatial_compare_plot_lb}).  We also see a distance modulus gradient of $0.099^{+0.014}_{-0.015}$ mag deg$^{-1}$, slightly lower than but comparable to the value measured by \citet{StreamsDESShipp2018}.  Overall, the two tails are similar, the density of each smoothly decreasing while the width slowly increases as you move further from Tucana III (Figure \ref{fig:des_tucana_iii_fit_params}).  We see one notable increase in width around $\phi_1 = -1$, where we observe a central component of the stream with a similar width to the rest of the stream and a diffuse over-density around it driving up the width (Figure \ref{fig:des_tucana_iii_spatial_plots}).

\subsubsection{Turranburra Stream}

\begin{figure}
    \centering
    \includegraphics[width=\columnwidth]{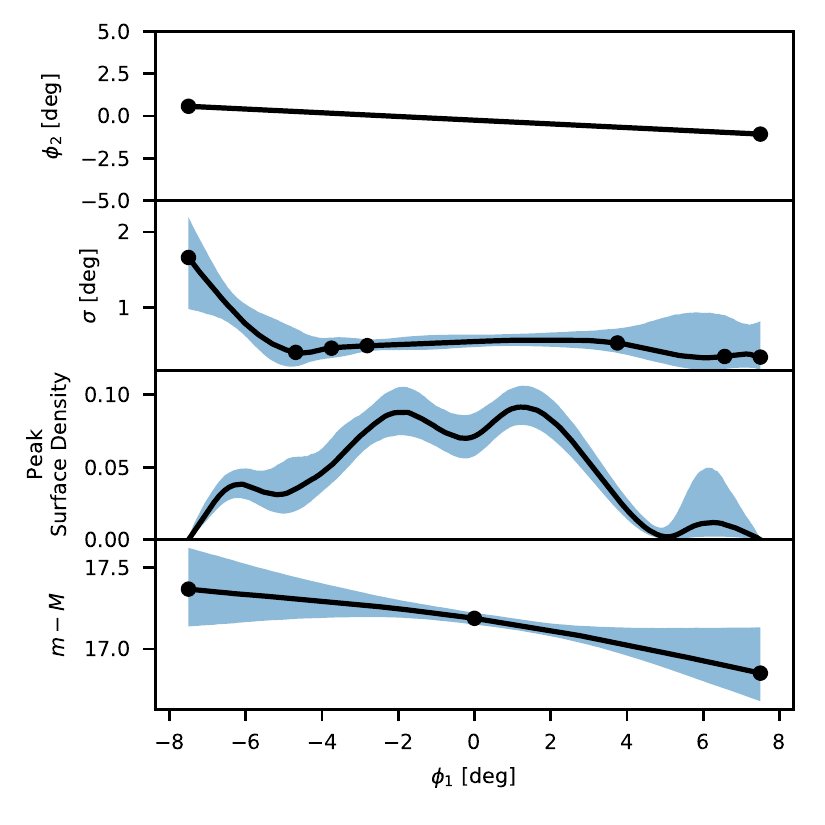}
    \caption{Summary of stream properties for Turranburra from our best fit model, with median (black) and 16th/84th percentiles (shaded blue) curves.  Black points are the final node positions for our model.}
    \label{fig:des_turranburra_fit_params}
\end{figure}

\begin{figure}
    \centering
    \includegraphics[width=\columnwidth]{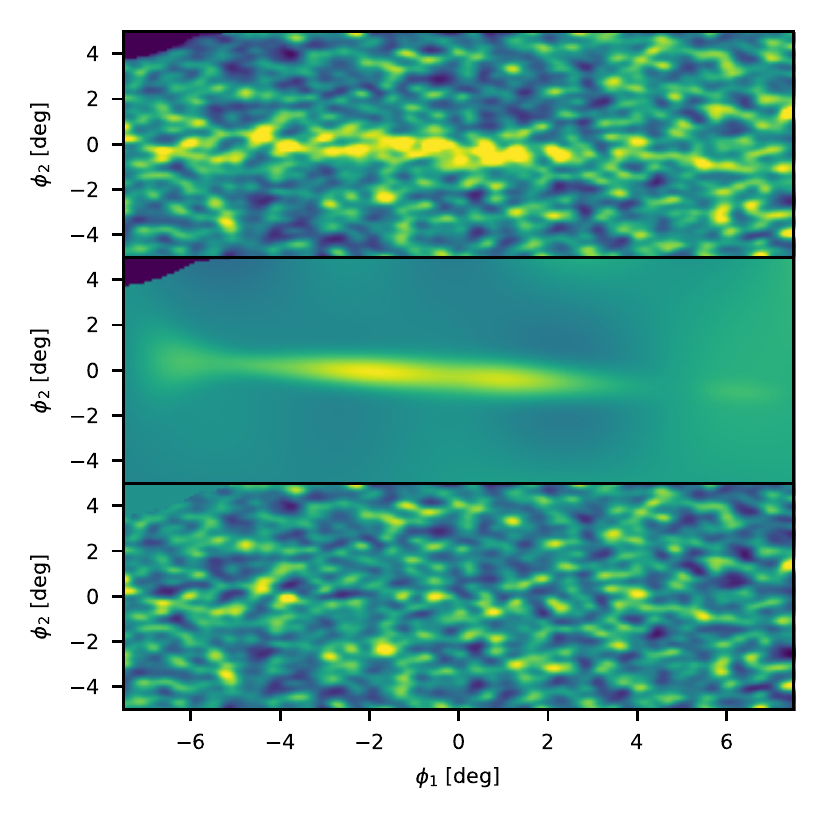}
    \caption{Top: Number density of DES stars near Turranburra with a Gaussian filter with standard deviation of 0.15 degrees applied.  Stars were selected using an matched filter based on a stellar population with $log_{10}(age) = 10.1$, [Fe/H] = -1.6, and distance modulus of 17.18.  Middle:  Best fit model of the stream, normalised to match the number of stars in the top plot.  Bottom: Residual stellar density with colour range centered at 0.}
    \label{fig:des_turranburra_spatial_plots}
\end{figure}

Another stream discovered in \citet{StreamsDESShipp2018}, Turranburra is a low density stream near the edge of the DES footprint.  Similar to \citet{StreamsDESShipp2018}, we are uncertain of the full extent of the stream, and we choose to focus on fitting a shorter 8 degree (8.2 kpc) segment of the stream away from the edge of the DES footprint.  The results of our fit are shown in Figures \ref{fig:des_turranburra_fit_params} and \ref{fig:des_turranburra_spatial_plots}.  For this segment, we see a distance modulus gradient along its length of $-0.029^{+0.019}_{-0.021}$ mag deg$^{-1}$ and minimal variation in width and density (Figure \ref{fig:des_turranburra_fit_params}).  At the southern tip of the stream, the width narrows slightly before an abrupt end of the stream at $\phi_1 = 4$, while the northern side of the stream sees the stream potentially continue with much lower surface density and larger width (Figure \ref{fig:des_turranburra_spatial_plots}).

\subsubsection{Willka Yaku Stream}

\begin{figure}
    \centering
    \includegraphics[width=\columnwidth]{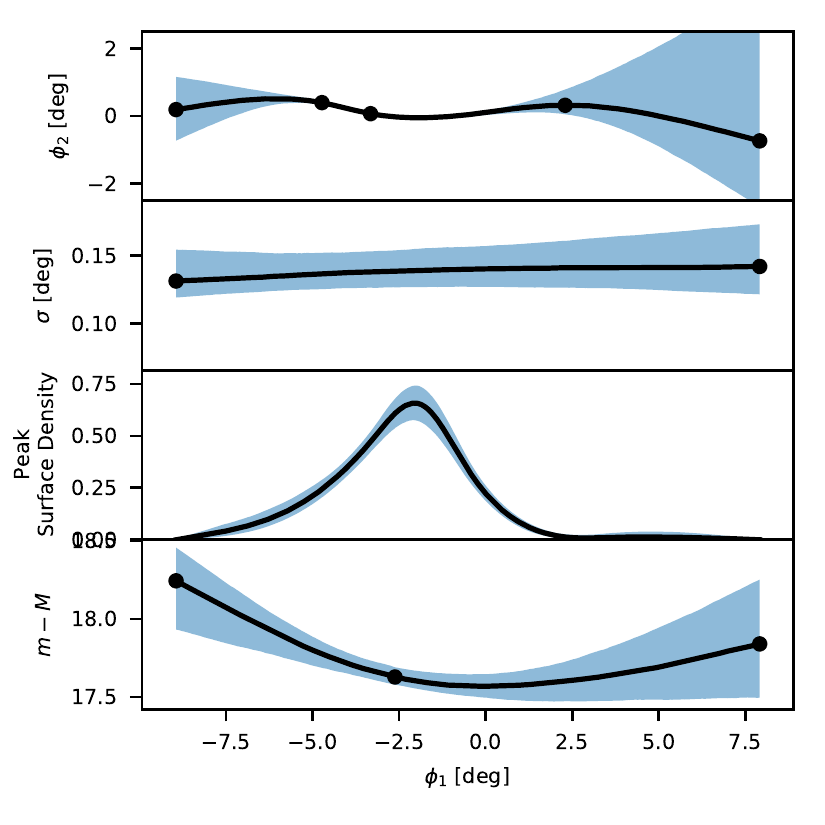}
    \caption{Summary of stream properties for Willka Yaku from our best fit model, with median (black) and 16th/84th percentiles (shaded blue) curves.  Black points are the final node positions for our model.}
    \label{fig:des_willka_yaku_fit_params}
\end{figure}

\begin{figure}
    \centering
    \includegraphics[width=\columnwidth]{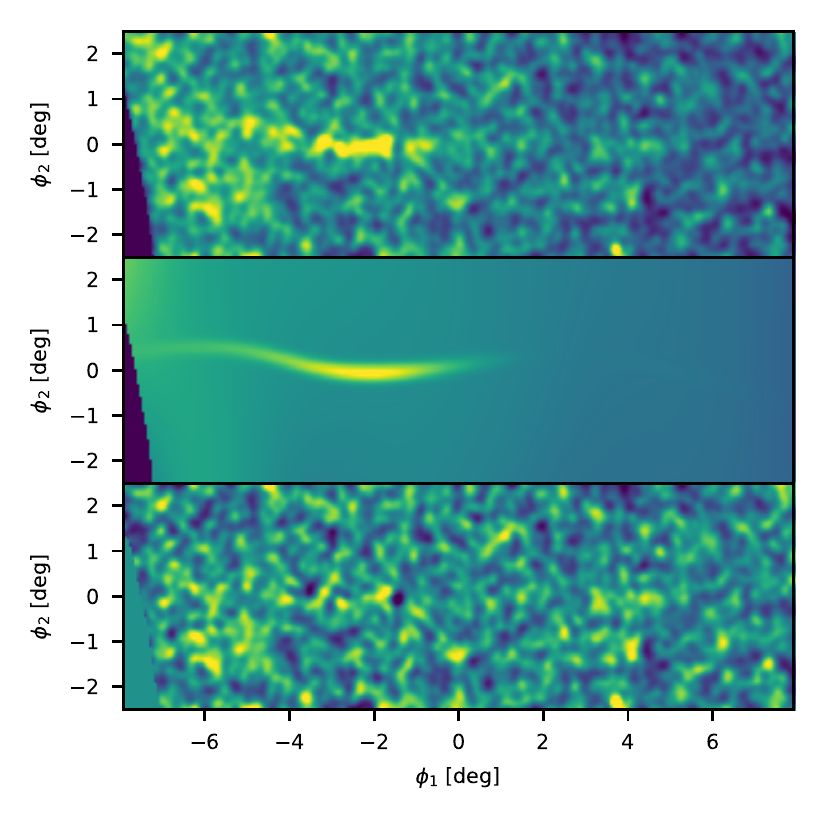}
    \caption{Top: Number density of DES stars near Willka Yaku with a Gaussian filter with standard deviation of 0.1 degrees applied.  Stars were selected using an matched filter based on a stellar population with $log_{10}(age) = 10.0$, [Fe/H] = -0.8, and distance modulus of 17.96.  Middle:  Best fit model of the stream, normalised to match the number of stars in the top plot.  Bottom: Residual stellar density with colour range centered at 0.}
    \label{fig:des_willka_yaku_spatial_plots}
\end{figure}

\citet{StreamsDESShipp2018} discovered Willka Yaku at the southern edge of DES at a distance around 37.7 kpc.  The results of our analysis are shown in Figures \ref{fig:des_willka_yaku_fit_params} and \ref{fig:des_willka_yaku_spatial_plots}.  We are limited by the edge of the dataset in the lower left of Figure \ref{fig:des_willka_yaku_spatial_plots} where there is no DES data.  We were still able to fit the full extent of the stream, spanning 7 degrees (8.3 kpc) with an average width of 0.14 degrees (83 pc).  The stream track has a slight curvature, and there is a small distance modulus gradient for $\phi_1 < 0$ of $-0.061^{+0.034}_{-0.024}$ mag deg$^{-1}$ (Figure \ref{fig:des_willka_yaku_fit_params}).  We begin losing the stream for $\phi_1 > 0$ and the distance modulus uncertainty increases significantly.  We see a bright, central region spanning 2 degrees from $\phi_1$ = -1.75 to -3.75 and peaking at $\phi_1 = -2$, with lower density tails extending further approximately 1 to 1.5 degrees North and 2 to 2.5 degrees South.  We smooth over an over-density at $\phi_1 = -1.75$ and a potential gap at about $\phi_1 = -1.5$ visible in Figure \ref{fig:des_willka_yaku_spatial_plots}.  The small sizes and close proximity of the two features ($\sim 0.1$ degree) cannot be resolved with our initial density node spacing of 1.5 degrees and node placement algorithm.

\subsection{Streams in DECaLS}

\subsubsection{GD-1 Stream}

\begin{figure}
    \centering
    \includegraphics[width=\columnwidth]{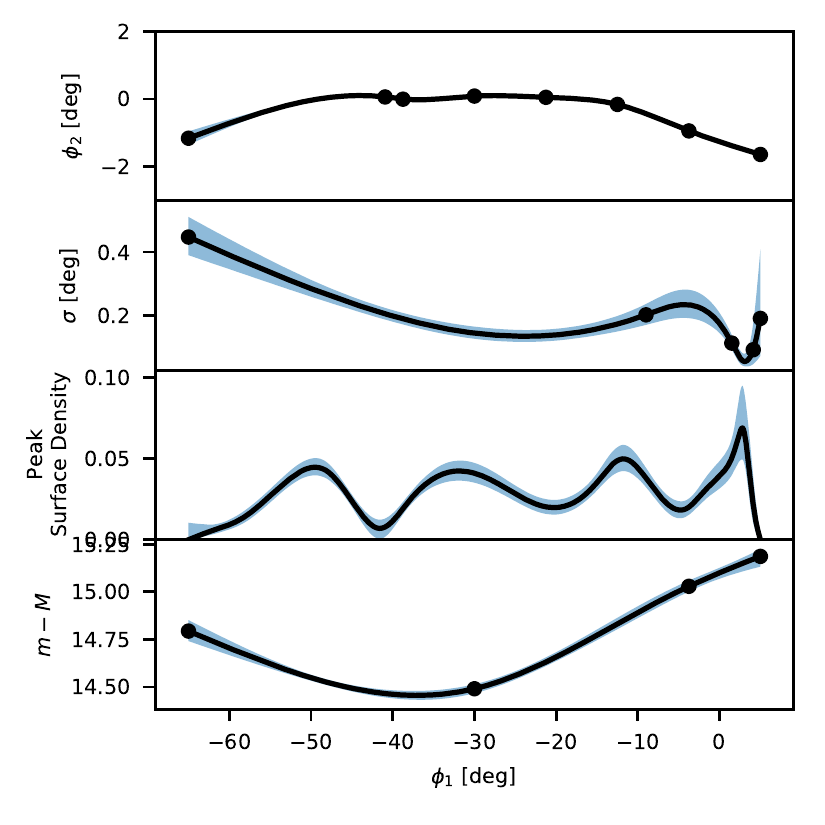}
    \caption{Summary of stream properties for GD-1 from our best fit model, with median (black) and 16th/84th percentiles (shaded blue) curves.  Black points are the final node positions for our model.}
    \label{fig:decals_gd_1_fit_params}
\end{figure}

\begin{figure}
    \centering
    \includegraphics[width=\columnwidth]{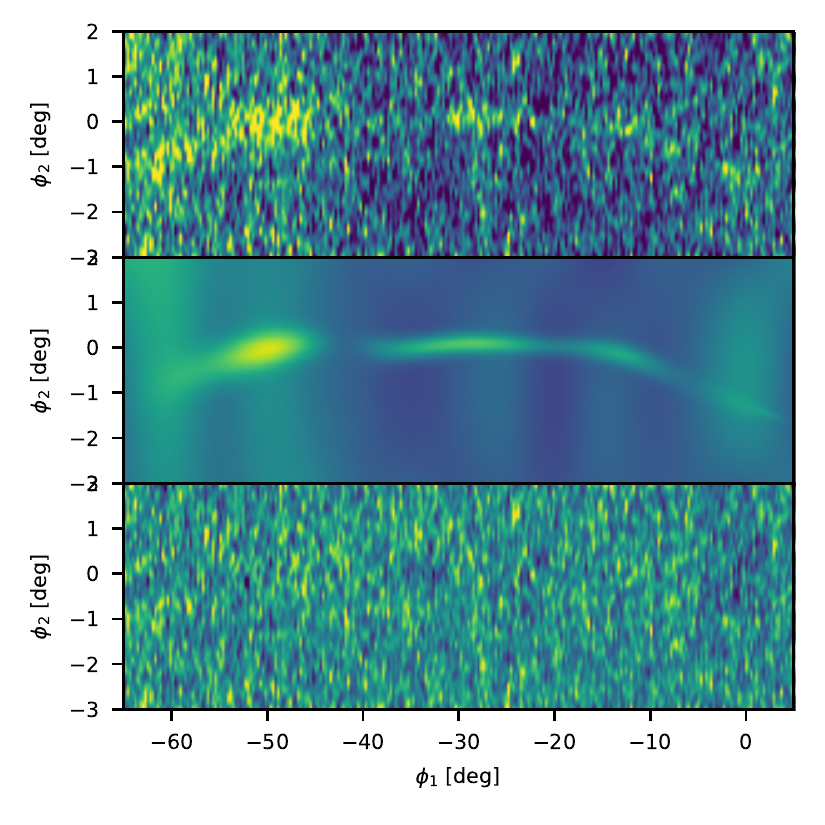}
    \caption{Top: Number density of DECaLS stars near GD-1 with a Gaussian filter with standard deviation of 0.1 degrees applied.  Stars were selected using an matched filter based on a stellar population with $log_{10}(age) = 10.05$, [Fe/H] = -1.2, and distance modulus of 14.6.  Middle:  Best fit model of the stream, normalised to match the number of stars in the top plot.  Bottom: Residual stellar density with colour range centered at 0.}
    \label{fig:decals_gd_1_spatial_plots_14_6}
\end{figure}

\begin{figure}
    \centering
    \includegraphics[width=\columnwidth]{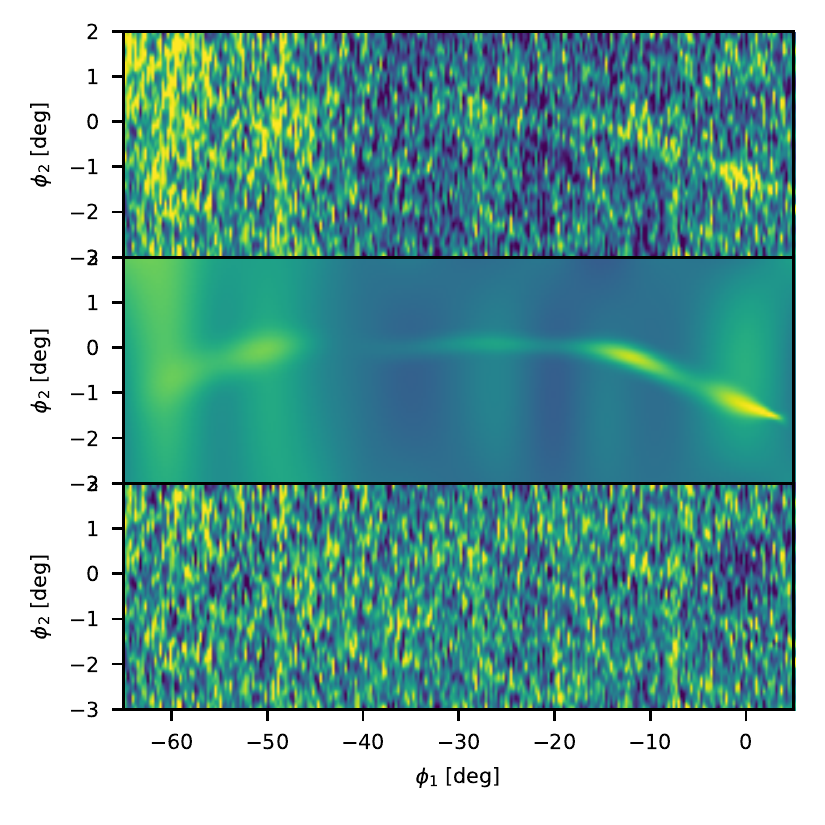}
    \caption{Same plots as Figure \ref{fig:decals_gd_1_spatial_plots_14_6} but using a distance modulus of 14.9.}
    \label{fig:decals_gd_1_spatial_plots_14_9}
\end{figure}

\begin{figure}
    \centering
    \includegraphics[width=\columnwidth]{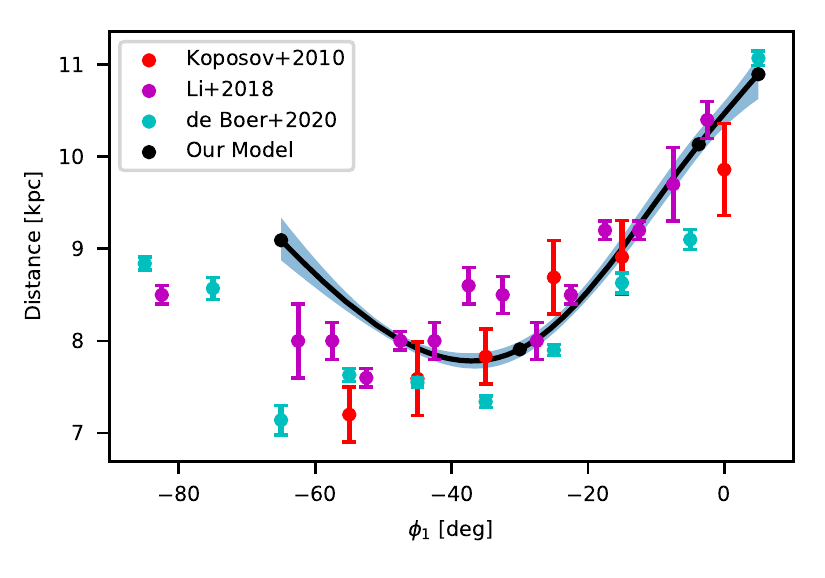}
    \caption{Plot of distance versus $\phi_1$ position for GD-1 stream comparing our model (black/blue) to measurements by \citet{ConstrainingGD1Koposov2010} (red), \citet{GD_1_relic_Li2018a} (magenta), and \citet{closer_look_gd_1_de_Boer_2020} (cyan)}
    \label{fig:decals_gd_1_dist_mod_compare}
\end{figure}

GD-1 is one of the longest streams, spanning over 90 degrees across the Northern sky \citep{GD_1_relic_Li2018a}.  First discovered by \citet{GD_1_discovery_Grillmair_2006}, GD-1 has been used extensively to constrain the gravitational potential of the Milky Way \citep{ConstrainingGD1Koposov2010, Dipping_GD_1_Bowden_2015, Shape_Pal_5_GD_1_Bovy_2016}.  The results of our analysis are shown in Figures \ref{fig:decals_gd_1_fit_params}, \ref{fig:decals_gd_1_spatial_plots_14_6}, and \ref{fig:decals_gd_1_spatial_plots_14_9}.  Due to its length, it begins to intersect the Sagitarrius stream and galactic plane and must limit our region of interest to $\phi_1 > -65$.  Within this range, we fit a total length of 65 degrees (11 kpc) with an average width of 0.2 degrees (30 pc).  For $\phi_1 > -40$, the width of the stream remains fairly constant (Figure \ref{fig:decals_gd_1_spatial_plots_14_9}), but in Figure \ref{fig:decals_gd_1_spatial_plots_14_6} we can see a notable increase in width for $\phi_1 < -40$.  GD-1 also has a significant distance modulus gradient and curvature, with the closest distance to the sun at $\phi_1 = -35$ (Figure \ref{fig:decals_gd_1_fit_params}).  This can be seen comparing Figures \ref{fig:decals_gd_1_spatial_plots_14_6} and \ref{fig:decals_gd_1_spatial_plots_14_9}, created using a matched filter with a distance modulus of 14.6 and 14.9 mag respectively.  We see a distance modulus gradient of $0.017^{+0.001}_{-0.001}$ mag deg$^{-1}$ for $\phi_1 > -37$ and $-0.011^{+0.002}_{-0.002}$ mag deg$^{-1}$ for $\phi_1 < -37$.  The distance gradient is similar to \citet{ConstrainingGD1Koposov2010}, \citet{GD_1_relic_Li2018a}, and \citet{closer_look_gd_1_de_Boer_2020} above $\phi_1 > -37$ and deviates slightly for $\phi_1 < -37$ (Figure \ref{fig:decals_gd_1_dist_mod_compare}).  Our model fits a more symmetric gradient around the point of closest approach, compared to the more gradual slope for $\phi_1 > -35$ seen by \citet{GD_1_relic_Li2018a}.

We find several gaps and density variations in GD-1 along its length, with a gap at $\phi_1 = -40$ and two under-densities at $\phi_1 = -20$ and $-4$.  We see a consistent peak surface brightness between the gaps, with a higher linear density for $\phi_1 < -40$ offset by the increase in width of the stream.  The locations of these under-densities and overall density profile is in agreement with past papers on GD-1, except for the depth of the gap at $\phi_1 = -40$ where we see a more pronounced decrease in stream density \citep{Spur_Gap_GD_1_Bonaca2018, GD_1_relic_Li2018a, closer_look_gd_1_de_Boer_2020}.  However, GD-1 has a spur around $\phi_1 = -40$ \citep{Spur_Gap_GD_1_Bonaca2018} that we are unable to include in our model, since we cannot fit structures off the main stream track.  The spur would instead be included in the background model and reduce the density of our stream model in that area, contributing to the difference seen between our stream model and the other papers.  This effect is similar to our limitations fitting the narrow component of Jhelum.

\subsubsection{Palomar 5}

One of the oldest known streams, the tidal tails of Palomar 5 were first discovered by \citet{Pal_5_Tidal_Discovery_Odenkirchen_2001} using data from SDSS.  They initially detected short tails on either side of the Palomar 5 cluster, extending only a few degrees on the sky.  However, later studies on Pal 5 showed that the stream is not only significantly longer, covering tens of degrees, but it also contains a variety of substructure \citep{Pal_5_extended_tail_Grillmair2006, Pal_5_gaps_carlberg_2012,  Pal_5_Variations_Bonaca_2020}.  The stream track on the sky shows as strong S-shape behaviour with a significant shift between the leading and trailing tails near the progenitor \citep{SharperViewErkal2017}.  Because of this shift, we fit the leading and trailing tails separately and mask the Pal 5 cluster itself.  We also mask the M5 cluster near $(\phi_1, \phi_2) = (2, 1)$.  We use the stream aligned coordinate system defined in \citet{SharperViewErkal2017}, which places the Pal 5 cluster at $(\phi_1, \phi_2) = (-0.07, -0.13)$.  We lack data in the lower left and right of each tidal tail's region due to limits in DECaLS spatial coverage. 

\begin{figure}
    \centering
    \includegraphics[width=\columnwidth]{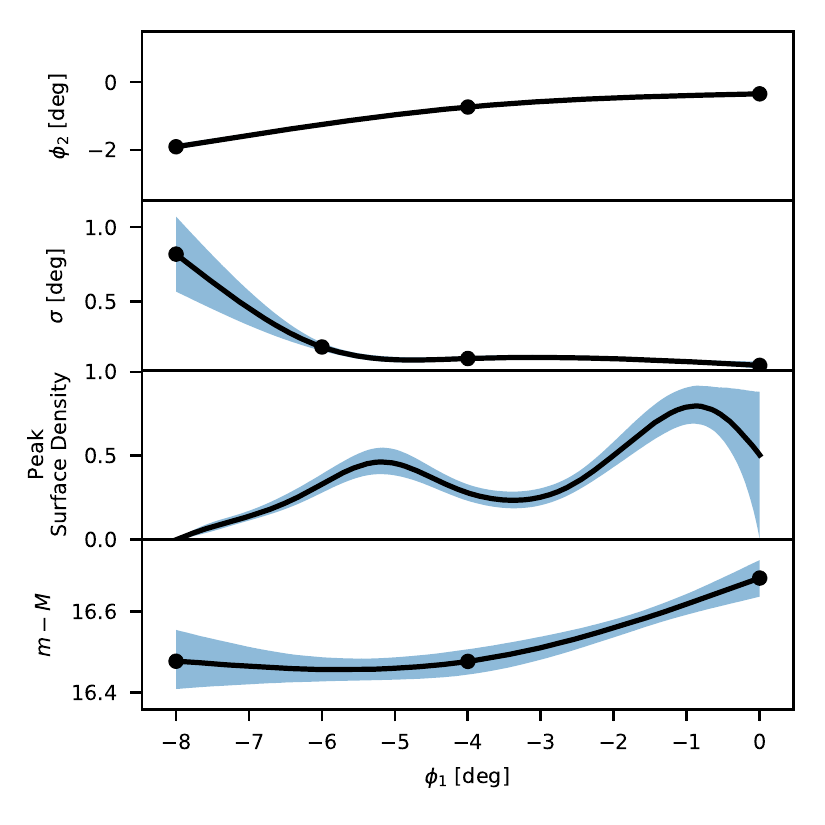}
    \caption{Summary of stream properties for the leading tail of Palomar 5 from our best fit model, with median (black) and 16th/84th percentiles (shaded blue) curves.  Black points are the final node positions for our model.}
    \label{fig:decals_pal_5_leading_fit_params}
\end{figure}

\begin{figure}
    \centering
    \includegraphics[width=\columnwidth]{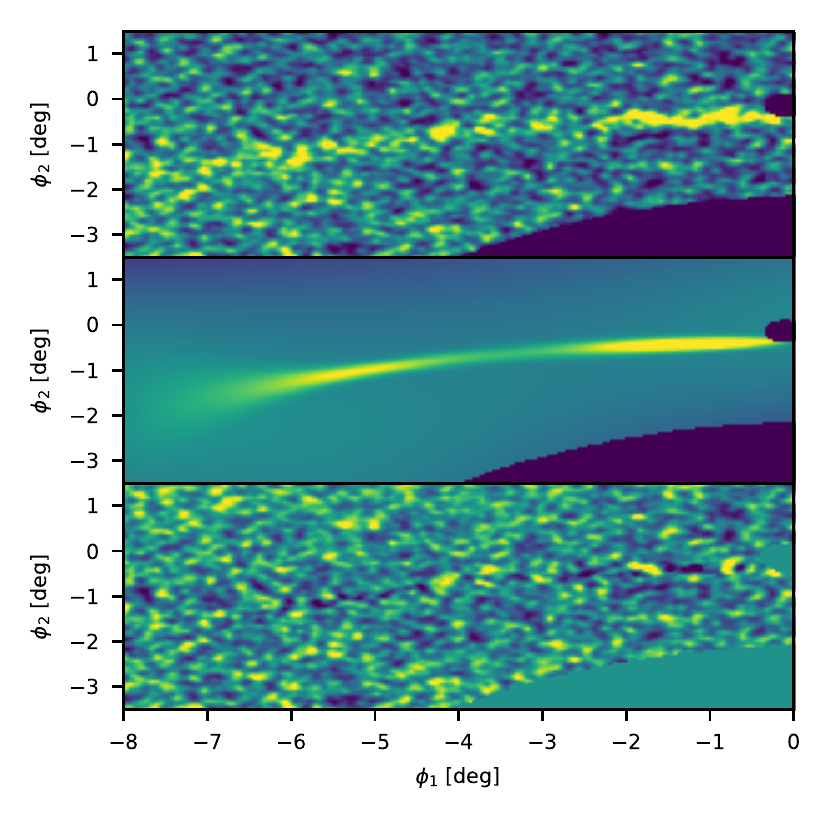}
    \caption{Top: Number density of DECaLS stars near the leading tail of Palomar 5 with a Gaussian filter with standard deviation of 0.05 degrees applied.  Stars were selected using an matched filter based on a stellar population with $log_{10}(age) = 10.05$, [Fe/H] = -1.1, and distance modulus of 16.66.  Middle:  Best fit model of the stream, normalised to match the number of stars in the top plot.  Bottom: Residual stellar density with colour range centered at 0.}
    \label{fig:decals_pal_5_leading_spatial_plots}
\end{figure}

\begin{figure}
    \centering
    \includegraphics[width=\columnwidth]{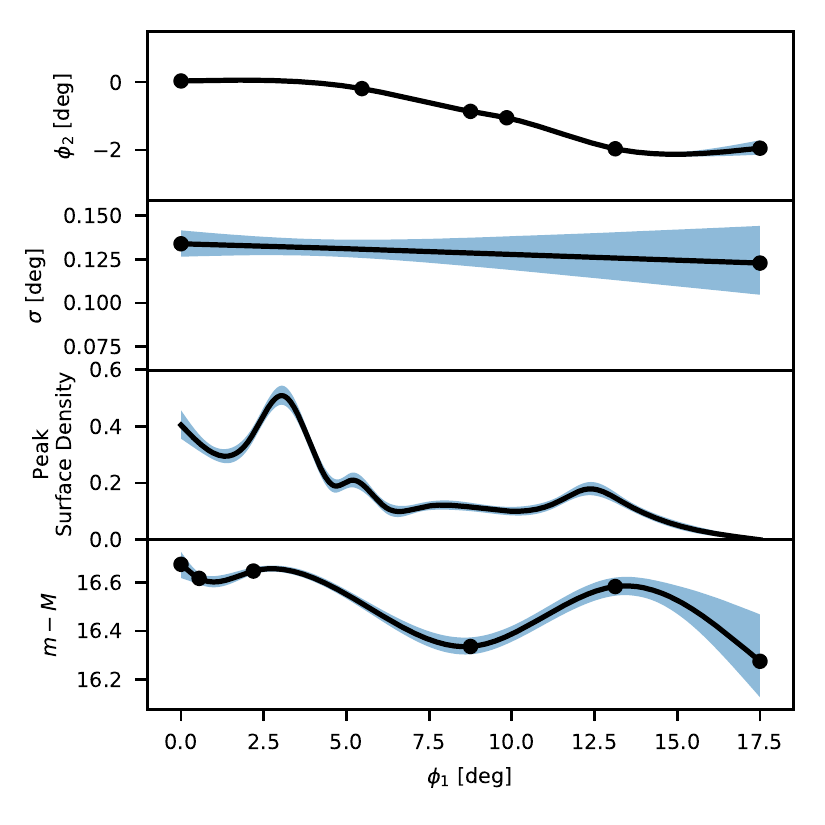}
    \caption{Summary of stream properties for the trailing tail of Palomar 5 from our best fit model, with median (black) and 16th/84th percentiles (shaded blue) curves.  Black points are the final node positions for our model.}
    \label{fig:decals_pal_5_trailing_fit_params}
\end{figure}

\begin{figure}
    \centering
    \includegraphics[width=\columnwidth]{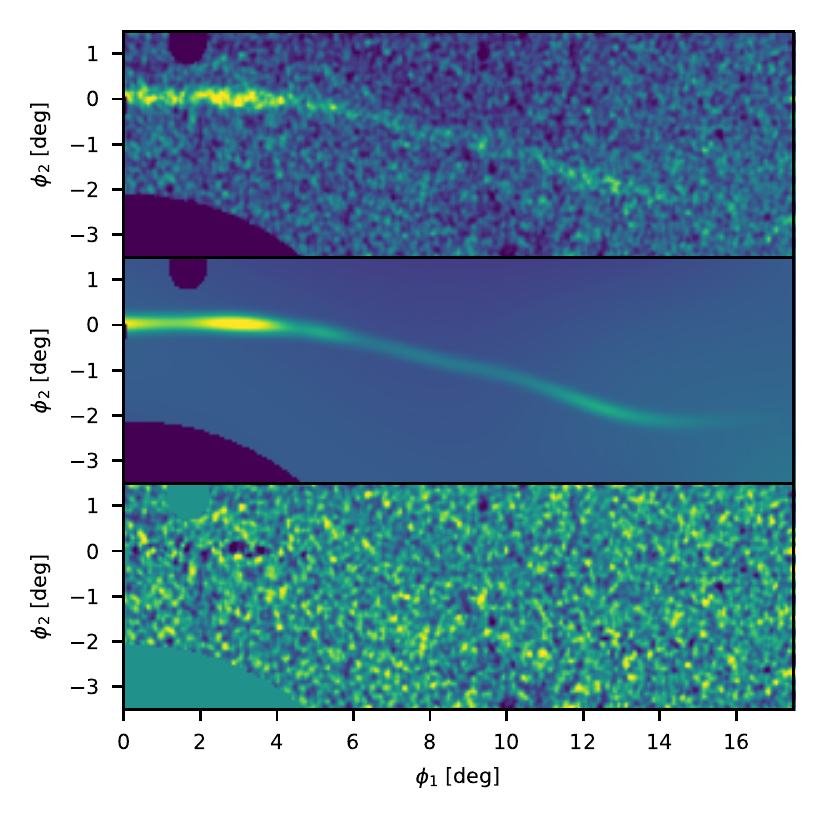}
    \caption{Top: Number density of DECaLS stars near the trailing tail of Palomar 5 with a Gaussian filter with standard deviation of 0.05 degrees applied.  Stars were selected using an matched filter based on a stellar population with $log_{10}(age) = 10.05$, [Fe/H] = -1.1, and distance modulus of 16.66.  Middle:  Best fit model of the stream, normalised to match the number of stars in the top plot.  Bottom: Residual stellar density with colour range centered at 0.}
    \label{fig:decals_pal_5_trailing_spatial_plots}
\end{figure}

Overall, our results agrees closely with \citet{SharperViewErkal2017} with some slight differences in stream density.  The results of our analysis on the leading tail are shown in Figures \ref{fig:decals_pal_5_leading_fit_params} and \ref{fig:decals_pal_5_leading_spatial_plots} while the trailing tail results are shown in Figures \ref{fig:decals_pal_5_trailing_fit_params} and \ref{fig:decals_pal_5_trailing_spatial_plots}.  Given the prominence of Palomar 5 and its tidal tails, we are unsurprised our stream tracks match \citet{SharperViewErkal2017}.  We see a stable width of around 0.12 degrees along the majority of the trailing tail (Figure \ref{fig:decals_pal_5_trailing_fit_params}) but slight broadening from 0.1 degrees to 0.2 degrees as the leading tail extends away from the cluster (Figure \ref{fig:decals_pal_5_leading_fit_params}).  We also see the same "fanning" structure noted by \citet{Pal_5_Variations_Bonaca_2020} past $\phi_1 > -6$, where the leading tail begins to broaden significantly more as we reach the end of the stream.  Overall, we are able to fit a total length of 22.5 degrees, including 7.5 degrees (3.9 kpc) for the leading tail and 15 degrees (9.4 kpc) for the trailing tail.  There are slight differences when looking at the stream's stellar density along its length compared to \citet{SharperViewErkal2017} primarily due to differences in model design.  We use a coarser starting grid of density nodes and look to prune those nodes first, limiting the size of features our model will resolve.  As a result, we smooth over density perturbations in the 2-3 degrees on either side of the cluster compared to the narrow features \citet{SharperViewErkal2017} highlights at $\phi_2$ = -2, -0.75, and 0.75 (Figures \ref{fig:decals_pal_5_leading_spatial_plots} and \ref{fig:decals_pal_5_trailing_spatial_plots}).  However, we still detect the narrow over-density at $\phi_1 = -0.89$ visible in Figure \ref{fig:decals_pal_5_leading_spatial_plots} and two major under-densities at $\phi_1 = -3.4$ and from $\phi_1$ = 6 to 10 that could possibly be explained by interaction with dark matter sub-halos \citep{SharperViewErkal2017, Pal_5_Variations_Bonaca_2020}.

\subsubsection{Triangulum Stream}

\begin{figure}
    \centering
    \includegraphics[width=\columnwidth]{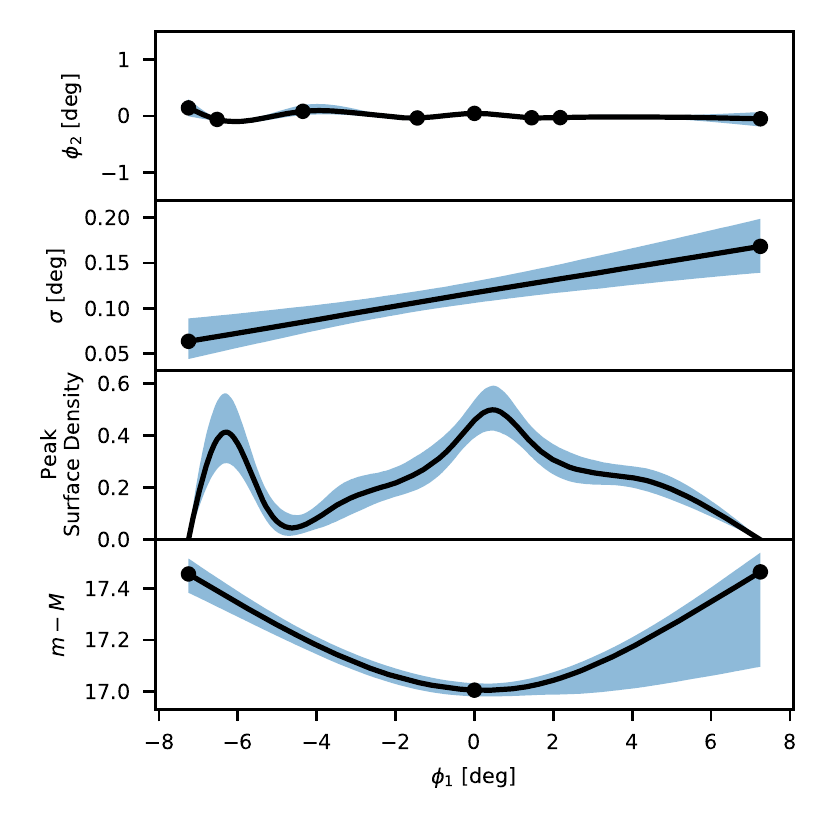}
    \caption{Summary of stream properties for Triangulum from our best fit model, with median (black) and 16th/84th percentiles (shaded blue) curves.  Black points are the final node positions for our model.}
    \label{fig:decals_triangulum_fit_params}
\end{figure}

\begin{figure}
    \centering
    \includegraphics[width=\columnwidth]{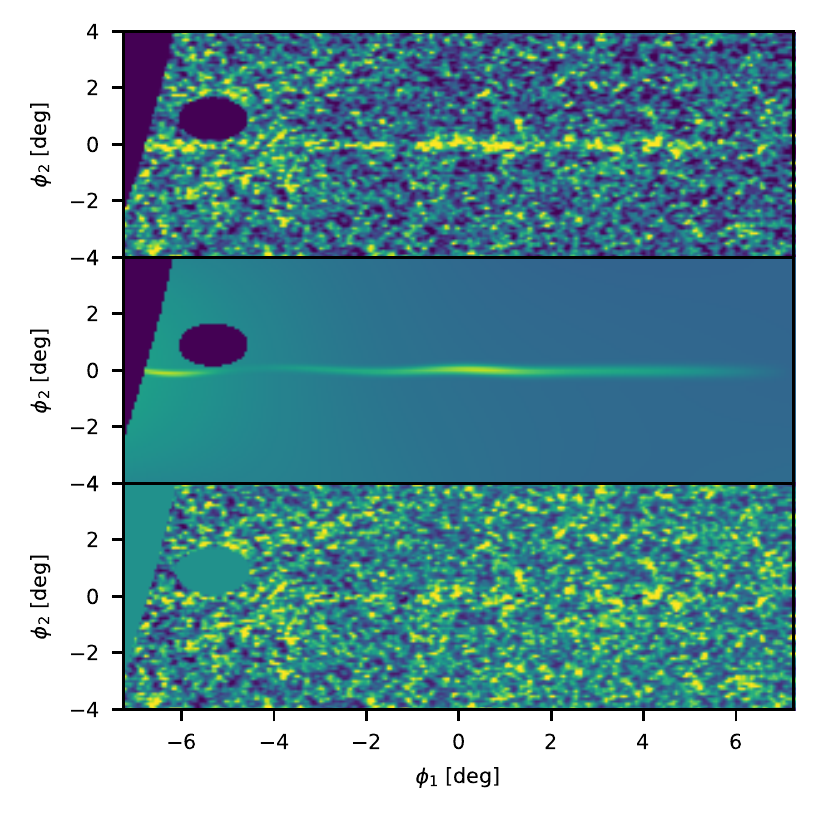}
    \caption{Top: Number density of DECaLS stars near Triangulum with a Gaussian filter with standard deviation of 0.05 degrees applied.  Stars were selected using an matched filter based on a stellar population with $log_{10}(age) = 10.05$, [Fe/H] = -1.1, and distance modulus of 16.96.  Middle:  Best fit model of the stream, normalised to match the number of stars in the top plot.  Bottom: Residual stellar density with colour range centered at 0.}
    \label{fig:decals_triangulum_spatial_plots}
\end{figure}

The Triangulum stellar stream, also known as the Pisces stellar stream, is a thin stream at the far North of the SDSS footprint.  Triangulum was first discovered using SDSS photometry \citep{TriangulumBonaca2012} and confirmed kinematically using SDSS spectroscopy of red giant branch stars \citep{TriangulumKinematicsMartin2013}.  Similar to \citet{TriangulumBonaca2012} and \citet{TriangulumKinematicsMartin2013} with SDSS, we are limited by the edge of the DECaLS footprint at the top right of Figure \ref{fig:decals_triangulum_spatial_plots}. Additionally, we mask nearby M33 galaxy at $(\phi_1, \phi_2) = (-5, 1)$ to prevent it from interfering with the fit.  The results of our fit are shown in Figures \ref{fig:decals_triangulum_fit_params} and \ref{fig:decals_triangulum_spatial_plots}.  We are able to see a large extent of the stream, covering 13.5 degrees (12 kpc) with an average width of 0.12 degrees (54 pc).  Because we are able to detect a larger extent of the stream comparing to \citet{TriangulumKinematicsMartin2013}, we detect a change in distance along the length of the stream.  The stream is closest to the Sun at $\phi_1 = 0$, with an average distance modulus gradient of $-0.059^{+0.010}_{-0.009}$ mag deg$^{-1}$ and $0.062^{+0.011}_{-0.051}$ mag deg$^{-1}$ for $\phi_1 < 0$ and $> 0$ respectively (Figure \ref{fig:decals_triangulum_fit_params}).  While \citet{TriangulumKinematicsMartin2013} did not detect any distance gradient, we believe that is most likely due to their data consisting of a much shorter stream segment (6 vs 14 degrees for our data) in addition to their on stream region being centered on the point of closest approach.  We also observe two distinct segments of the stream, with a gap at approximately $\phi_1 = -4.5$ separating them.  While minor, we also note a slight disturbance to the stream track around the gap visible in Figure \ref{fig:decals_triangulum_fit_params}.  The $\phi_2$ position of the stream appear to oscillate by about $\pm0.1$ degrees from $\phi_1 = -7$ to $\phi_1 = 1.5$.

\subsection{Streams in Pan-STARRS}

\subsubsection{Ophiuchus Stream}

\begin{figure}
    \centering
    \includegraphics[width=\columnwidth]{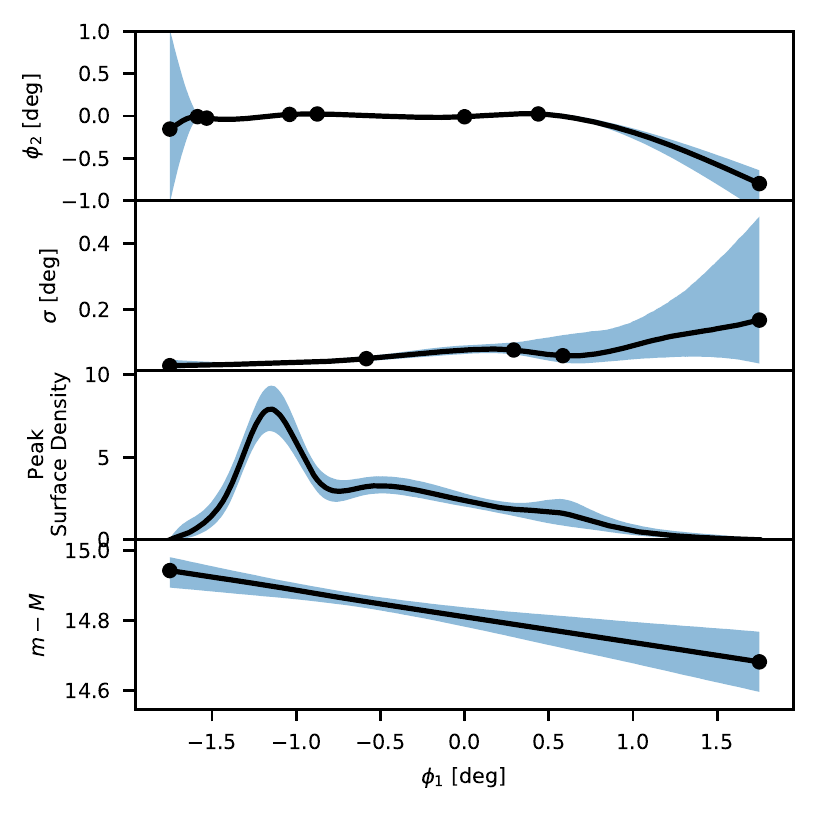}
    \caption{Summary of stream properties for Ophiuchus from our best fit model, with median (black) and 16th/84th percentiles (shaded blue) curves.  Black points are the final node positions for our model.}
    \label{fig:panstarrs_ophiuchus_fit_params}
\end{figure}

\begin{figure}
    \centering
    \includegraphics[width=\columnwidth]{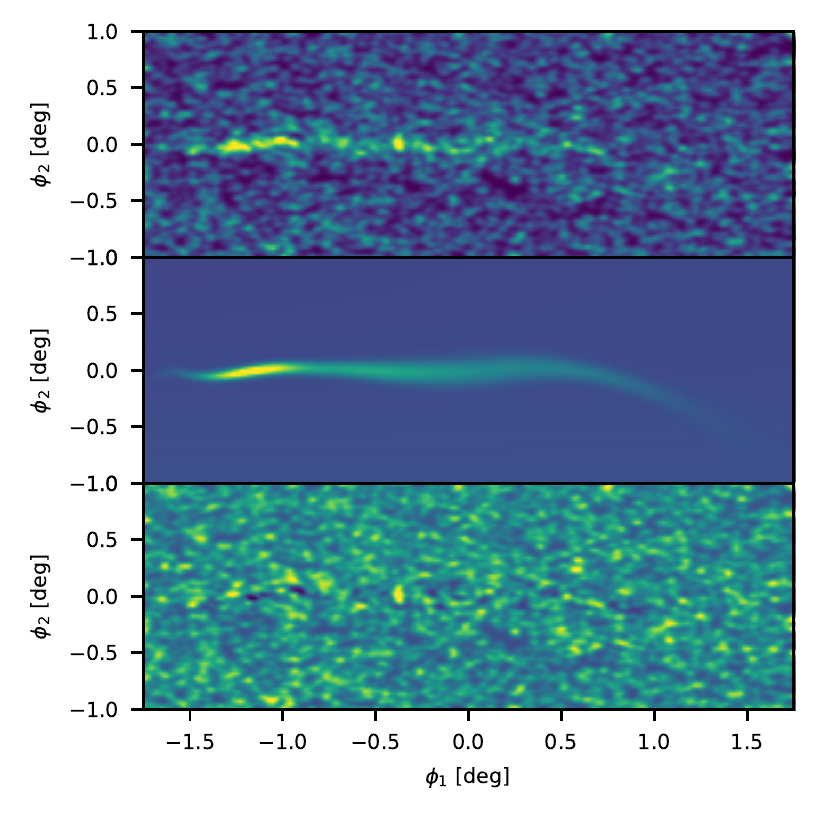}
    \caption{Top: Number density of Pan-STARRS stars near Ophiuchus with a Gaussian filter with standard deviation of 0.02 degrees applied.  Stars were selected using an matched filter based on a stellar population with $log_{10}(age) = 10.0$, [Fe/H] = -1.9, and distance modulus of 14.8.  Middle:  Best fit model of the stream, normalised to match the number of stars in the top plot.  Bottom: Residual stellar density with colour range centered at 0.}
    \label{fig:panstarrs_ophiuchus_spatial_plots}
\end{figure}

\begin{figure}
    \centering
    \includegraphics[width=\columnwidth]{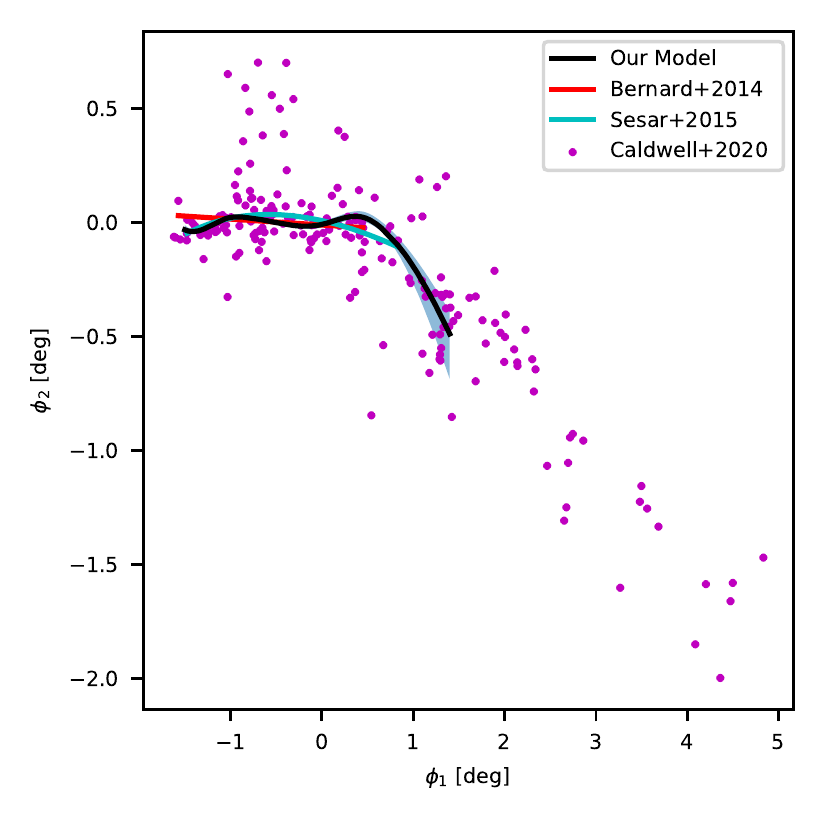}
    \caption{Our model of the Ophiuchus stream track (black, with 1-sigma uncertainties in blue) compared to measurements of the stream track by \citet{SerendipitousBernard_2014} (red curve), \citet{nat_orbit_Sesar_2015} (cyan curve). 
    Magenta points show likely Ophiuchus member stars from \citet{larger_extent_Caldwell_2020}.}
    \label{fig:panstarrs_ophiuchus_stream_track}
\end{figure}

The smallest stream we include in this paper, Ophiuchus was first discovered by \citet{SerendipitousBernard_2014}.  Since then, there have been several follow-up studies and observations, including spectroscopic observations of member stars \citep{nat_orbit_Sesar_2015, larger_extent_Caldwell_2020}.  The results of our analysis of Ophiuchus are shown in Figures \ref{fig:panstarrs_ophiuchus_fit_params} and \ref{fig:panstarrs_ophiuchus_spatial_plots}.  The total length we are able to observe is 2.9 degrees (1.0 kpc) with an average width of 0.06 degrees (9 pc) and a slight distance modulus gradient of $-0.075^{+0.037}_{-0.034}$ mag deg$^{-1}$.  While there are no gaps detected in the stream, there is a significant decrease in density around $\phi_1 = -1$ as we transition from a large over-density to an extended, low density tail (Figure \ref{fig:panstarrs_ophiuchus_spatial_plots}).  From there, the stream density slowly decreases until we can no longer detect the stream around $\phi_1 = 1.5$.  We also see that the stream begin to curve around $\phi_1 = 1$ in Figure \ref{fig:panstarrs_ophiuchus_fit_params}.  This curve extends past where some previous photometric studies were able to detect, although more recent spectroscopic observations \citep{larger_extent_Caldwell_2020} have identified member stars continuing potentially twice our detected length (Figure \ref{fig:panstarrs_ophiuchus_stream_track}).

\subsection{Other Streams}
\label{sec:otherstreams}

While we were able to fit many Milky Way streams, we unfortunately could not include a number of known streams in this paper.  The majority of excluded streams had insufficient detected stream stars or surface density to be fit using our model, such as Styx \citep{Grillmair_Styx_2009}, Molonglo \citep{StreamsDESShipp2018}, Jet \citep{delve_jet_ferguson2021delveing}, and PS1B \citep{Panstarrs_Halo_Substructures_Bernard_2016}.  Without a sufficient number of stream stars, we failed to detect the stream during the initial fitting or fail to accurately model it with cubic splines. We also did not try to fit large streams, such as Sgr \citep{FieldofStreams_Belokurov_2006, Sagittarius_Koposov2012} or Palca \citep{StreamsDESShipp2018}, as those cover vast area of the sky and cross other Milky Way substructures that are not incorporated well into our single component background model, reducing the accuracy of the overall model. Similarly, we could not fit several smaller streams due to other large interfering structures passing directly through them, such as the Sgr stream passing through the Orphan stream, 300S, and PS1A \citep{Panstarrs_Halo_Substructures_Bernard_2016}.  Also all the streams discovered during the last two years, such as the tidal tails of Palomar 13 \citep{Shipp_Pal13_2020}, were not considered in this paper to avoid delaying the preparation of this paper.

\section{Discussion}
\label{sec:discussion}

In previous sections we demonstrated uniform modeling of stellar densities and color-magnitude diagrams of a large number of known stellar streams.  Using these models, we characterised global stream properties such as stream widths, distance moduli, total luminosities as well as identified a variety of substructures within the streams.  In case of several streams, such as AAU, GD-1, and Triangulum, they show stream gaps with stream density dropping almost to zero,  with two segments of the stream separated by as small as a tenth of a degree or up to almost 10 degrees. Some streams, like AAU, Elqui, Phoenix, and Ophiuchus, also show compact over-densities, where the density of the stream is two or three times higher the average density.  The width of most streams remained approximately constant along their length, with slight broadening occurring at their ends.  Streams with gaps, such as Aliqa Uma, Atlas, or GD-1, often show additional broadening where the stream gap occurred.  We also observed two stream tracks, for AAU and Triangulum, that significantly deviated from the simpler linear or curved tracks most streams had.  The stream track for AAU contains a large discontinuity, separating the Atlas and Aliqa Uma segments of the stream, while Triangulum exhibited noticeable wiggles in its stream track for approximately half its length.  All this shows that pretty much all streams are far from simple great circle arcs with monotonically changing densities. What is the cause of such complexity and substructure is I think still unclear. Multiple hypotheses have been brought up, interaction with dark matter subhalos \citep{SharperViewErkal2017, closer_look_gd_1_de_Boer_2020}, with the bar \citep{ophiuchus_galactic_bar_Price_Whelan_2016, Pal_5_bar_Pearson2017}, giant molecular clouds \citep{baryonic_DM_pal_5_Banik2019}, epiciclic overdensities 
\citep{epicyclic_OD_Tidal_Kupper_2010, feathers_bifurcations_shells_amorisco_2015, epicyclic_OD_Pal_5_Kupper_2015}, interaction with parent dwarf galaxies \citep{butterfly_cocoon_malhan_2019, closer_look_gd_1_de_Boer_2020}, or interactions with other massive bodies such as the LMC or Sgr dwarf galaxy \citep{brokenatlasaliqali2020, Jhelum_SGR_Dynamics_Woudenberg_2022}. Further observations are certainly needed to solve this.

While in this paper, we primarily focused on the individual stream measurements for each stream, in Section \ref{sec:progenitor_connections} we take a first look at a combined analysis of the streams as a population. We remark that since in this paper did not rely on any kinematic data, such as radial velocities or proper motions, our analysis is somewhat limited \citep[comparing to e.g.][]{chem_prop_twelve_streams_li2021s5}. We leave the analysis of our stream measurements in combination with {\it Gaia} and radial velocity data to a future contributions.

\subsection{Connections Between Stellar Streams Populations and Their Progenitors}
\label{sec:progenitor_connections}

\begin{figure}
    \centering
    \includegraphics[width=\columnwidth]{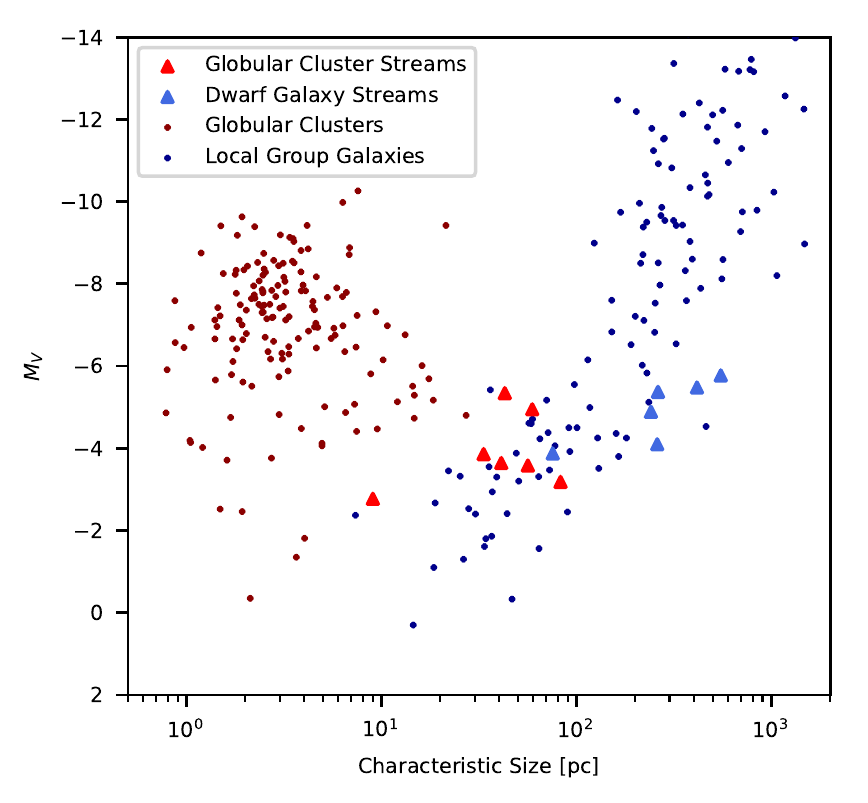}
    \caption{Absolute magnitude vs characteristic size for all fit stellar streams and known GCs and Local Group galaxies.  We use average width in parsecs as the characteristic size of a stellar stream, and half-light radius for GCs and galaxies.  Streams are classified as having either a globular cluster progenitor (red triangles) or dwarf galaxy progenitor (blue triangles) based on chemical analysis.  Known globular clusters (dark red circles) are from \citet{GC_catalog_Harris1996} (2010 Edition) and known Local Group galaxies (dark blue circles) are from \citet{dwarf_catalogue_mcconnachie_2012} (January 2021 Edition).}
    \label{fig:width_vs_m_v}
\end{figure}

Stellar streams fall into two populations based on their progenitor, globular clusters (GC) or dwarf galaxies \citep{Intro_to_Streams_Newberg_2016}.  The defining difference between these two types of progenitors is their dark matter content, with globular clusters lacking dark matter while dwarf galaxies are embedded within a dark matter halo \citep{gc_dwarf_galaxy_bimodal_Gilmore_2007, gc_galaxy_formation_Beasley_2020}.  Dwarf galaxies also have different stellar populations from globular clusters with significant metallicity spread and mass-metallicity relation \citep{GalaxyDefinedWillman2012, DwarfGalaxiesSimon2019}.  In order to classify the population of streams considered in this paper, we used the results from chemical analysis studies of the streams.  Seven of the streams we fit most likely originated from GCs \citep{TriangulumKinematicsMartin2013, nat_orbit_Sesar_2015, High_res_gd_1_Bonaca_2020, chem_prop_twelve_streams_li2021s5, brokenatlasaliqali2020}, and six likely originated from dwarf or ultra-faint dwarf galaxies \citep{Eight_Ultrafaint_Drlica-Wagner2015, Chemical_Tuc_III_Marshall_2019, Chemical_Abundances_Seven_Streams_Ji_2020, chem_prop_twelve_streams_li2021s5}.  In Figure \ref{fig:width_vs_m_v}, we plot the absolute magnitude ($M_V$) of the streams we analyzed versus their average width from our analysis, grouping them based on their progenitor type.  We also include intact globular clusters from \citet{GC_catalog_Harris1996} (2010 Edition) and Local Group galaxies from \citet{dwarf_catalogue_mcconnachie_2012} (January 2021 Edition) in Figure \ref{fig:width_vs_m_v}, using their half-light radii in place of stream width on the x-axis.

Historically, the width of a stream has been used as one of the primary methods to identify the type of progenitor, particularly in the absence of spectroscopy.  We can see from Figure \ref{fig:width_vs_m_v} that this remains a reasonable method of identification in most cases, as on average the GC streams we measured are narrower than the dwarf galaxy streams. However, classification based on width becomes difficult for streams around 100 pc wide, as streams originating from ultra-faint dwarf galaxies begin to overlap with GC streams.  A similar behaviour can be seen in Figure \ref{fig:width_vs_m_v} where faint ($M_V > -6$) intact GC and dwarf galaxies start to overlap in size at half-light radii between 10 and 30 pc.

When looking at the luminosities of the streams in Figure \ref{fig:width_vs_m_v}, we see that the luminosities of the population of GC streams  is noticeably lower than what we expect for intact globular clusters.  Figure \ref{fig:width_vs_m_v} shows that the GC streams that we modeled have an $M_V$ between -2 and -6.  In contrast, the majority of intact globular clusters have an $M_V$ between -6 and -9.  This difference could be explained by the fact that the stellar density of GCs determines how susceptible they are to tidal disruption, with less luminous and less dense GCs more likely to form streams \citep{feathers_bifurcations_shells_amorisco_2015}.  Alternatively it is possible that the stream luminosities that we measure are significantly underestimated because we are missing parts of stellar streams with significantly lower surface brightness.  That can be caused either because parts of streams have been significantly perturbed by DM subhalos \citep{SharperViewErkal2017, High_res_gd_1_Bonaca_2020} or possibly because low mass stars are preferentially stripped first from GCs resulting in tails with many stars below the depth limit of the survey.  This effect can cause the tidal tails of more massive and luminous GCs to be harder to detect compared to less luminous GCs, further biasing the average luminosity of our population of detected streams \citep{devil_in_the_tails_Balbinot_2018}.  Additionally, survey footprint boundaries limit the detected length of some streams and would also reduce the apparent stellar mass of the stream.  Figure \ref{fig:width_vs_m_v} also suggests that dwarf galaxy streams are fainter than the majority of intact dwarf galaxies, as the distribution in $M_V$ of dwarf galaxies in the sample of \citep{dwarf_catalogue_mcconnachie_2012} is pretty flat.  However this is most likely spurious due to our selection of dwarf galaxy streams and the current detectability limit of ultra-faint dwarf galaxies \citep{Eight_Ultrafaint_Drlica-Wagner2015}.  Larger, more luminous streams such as Sagittarius and Palca were excluded from our analysis because they cross numerous other objects that could not be effectively masked or included in our background model.  Similarly, we could only fit the Chenab portion of the Orphan-Chenab (OC) stream since Orphan crosses Sgr and the galactic plane.  This makes a high luminosity stream appear to be a lower luminosity stream, as the whole OC stream has an $M_V \sim -10.8$ \citep{ChenabOrphanKoposov_2019} while the Chenab segment of the stream we fit has $M_V = -5.5$.

\section{Conclusions}
\label{sec:conclusion}

In this paper, we described a flexible stream density model that could be broadly applied to many streams in a uniform, semi-automated manner. We then fit this model to a large group streams with data from DES, DECaLS, and Pan-STARRS to produce detailed density maps for each stream.  The stream models provided measurements of distances to the streams, stream tracks, and widths by describing them by cubic splines.  We then used these models to infer stream stellar masses and luminosities. 

Overall we observed significant complexity in stream  density maps.  Nearly every stream contained an under-density or compact over-density, with several streams having significant gaps along their stream track.  A variety of other substructure was also evident, spanning from multiple stream components within Jhelum stream, the discontinuity in the AAU stream track, and the stream track oscillations evident in Triangulum stream track.  Stream widths and distance gradients were more consistent, with no significant small scale variations.  Most streams had a constant width and only experienced slight broadening either near their ends or where there were gaps in the stream.  

With a uniform set of measurements for each stream, we took a preliminary look at a global properties of the stream population.  Unsurprisingly,  similar to the half-light radii of their intact counterparts, we saw that streams originating from globular clusters have smaller widths on average compared to dwarf galaxy streams.  However, in some cases width alone can not be used to classify a stream since both populations can produce streams with the width of around 100 pc.  We also observed that our population of GC streams is significantly less luminous than typical globular clusters, most likely due to the fact that streams represent lower-density and lower-luminosity clusters that are easier to disrupt.

There are multiple areas where the stream measurements presented in this paper can be applied.  Several studies in the past have used extracted stellar stream densities to study possible interactions between dark matter sub-halos and streams such as AAU \citep{brokenatlasaliqali2020}, Pal 5 \citep{SharperViewErkal2017}, and GD 1 \citep{Spur_Gap_GD_1_Bonaca2018}). This can now be expanded to a larger set of streams we present here. In addition, our results, in combination with kinematic measurements \citep[such as][]{chem_prop_twelve_streams_li2021s5}, provide data on stellar streams that are necessary to fit stellar stream disruption models  \citep{TucanaIIIErkal_2018, Ophiuchus_progenitor_Lane_2020}.  These stream disruption models can then be applied to large groups of streams to help constrain the Galactic potential as well characterize the impact of the Large Magellanic Clouds on the Milky Way\citep[i.e.][]{shipp2021measuring}.  Finally the stream tracks, widths, distances and density maps are essential for selecting targets for spectroscopic surveys and subsequent stream member identification.

\section*{Acknowledgements}

% Funding

JP thanks George and Marjorie Pake for their financial support through the Pake Fellowship.  MGW and SK acknowledge support from the National Science Foundation grants AST-1813881 and AST-1909584.  

% open access
For the purpose of open access, the author has applied a Creative Commons Attribution (CC BY) licence to any Author Accepted Manuscript version arising from this submission.

% DES

This project used public archival data from the Dark Energy Survey (DES). Funding for the DES Projects has been provided by the U.S. Department of Energy, the U.S. National Science Foundation, the Ministry of Science and Education of Spain, the Science and Technology FacilitiesCouncil of the United Kingdom, the Higher Education Funding Council for England, the National Center for Supercomputing Applications at the University of Illinois at Urbana-Champaign, the Kavli Institute of Cosmological Physics at the University of Chicago, the Center for Cosmology and Astro-Particle Physics at the Ohio State University, the Mitchell Institute for Fundamental Physics and Astronomy at Texas A\&M University, Financiadora de Estudos e Projetos, Funda{\c c}{\~a}o Carlos Chagas Filho de Amparo {\`a} Pesquisa do Estado do Rio de Janeiro, Conselho Nacional de Desenvolvimento Cient{\'i}fico e Tecnol{\'o}gico and the Minist{\'e}rio da Ci{\^e}ncia, Tecnologia e Inova{\c c}{\~a}o, the Deutsche Forschungsgemeinschaft, and the Collaborating Institutions in the Dark Energy Survey.

The Collaborating Institutions are Argonne National Laboratory, the University of California at Santa Cruz, the University of Cambridge, Centro de Investigaciones Energ{\'e}ticas, Medioambientales y Tecnol{\'o}gicas-Madrid, the University of Chicago, University College London, the DES-Brazil Consortium, the University of Edinburgh, the Eidgen{\"o}ssische Technische Hochschule (ETH) Z{\"u}rich,  Fermi National Accelerator Laboratory, the University of Illinois at Urbana-Champaign, the Institut de Ci{\`e}ncies de l'Espai (IEEC/CSIC), the Institut de F{\'i}sica d'Altes Energies, Lawrence Berkeley National Laboratory, the Ludwig-Maximilians Universit{\"a}t M{\"u}nchen and the associated Excellence Cluster Universe, the University of Michigan, the National Optical Astronomy Observatory, the University of Nottingham, The Ohio State University, the OzDES Membership Consortium, the University of Pennsylvania, the University of Portsmouth, SLAC National Accelerator Laboratory, Stanford University, the University of Sussex, and Texas A\&M University.

Based in part on observations at Cerro Tololo Inter-American Observatory, National Optical Astronomy Observatory, which is operated by the Association of Universities for Research in Astronomy (AURA) under a cooperative agreement with the National Science Foundation.

% DECALS

The Legacy Surveys consist of three individual and complementary projects: the Dark Energy Camera Legacy Survey (DECaLS; Proposal ID \# 2014B-0404; PIs: David Schlegel and Arjun Dey), the Beijing-Arizona Sky Survey (BASS; NOAO Prop. ID \# 2015A-0801; PIs: Zhou Xu and Xiaohui Fan), and the Mayall z-band Legacy Survey (MzLS; Prop. ID \# 2016A-0453; PI: Arjun Dey). DECaLS, BASS and MzLS together include data obtained, respectively, at the Blanco telescope, Cerro Tololo Inter-American Observatory, NSF's NOIRLab; the Bok telescope, Steward Observatory, University of Arizona; and the Mayall telescope, Kitt Peak National Observatory, NOIRLab. The Legacy Surveys project is honored to be permitted to conduct astronomical research on Iolkam Du'ag (Kitt Peak), a mountain with particular significance to the Tohono O'odham Nation.

NOIRLab is operated by the Association of Universities for Research in Astronomy (AURA) under a cooperative agreement with the National Science Foundation.

This project used data obtained with the Dark Energy Camera (DECam), which was constructed by the Dark Energy Survey (DES) collaboration. Funding for the DES Projects has been provided by the U.S. Department of Energy, the U.S. National Science Foundation, the Ministry of Science and Education of Spain, the Science and Technology Facilities Council of the United Kingdom, the Higher Education Funding Council for England, the National Center for Supercomputing Applications at the University of Illinois at Urbana-Champaign, the Kavli Institute of Cosmological Physics at the University of Chicago, Center for Cosmology and Astro-Particle Physics at the Ohio State University, the Mitchell Institute for Fundamental Physics and Astronomy at Texas A\&M University, Financiadora de Estudos e Projetos, Fundacao Carlos Chagas Filho de Amparo, Financiadora de Estudos e Projetos, Fundacao Carlos Chagas Filho de Amparo a Pesquisa do Estado do Rio de Janeiro, Conselho Nacional de Desenvolvimento Cientifico e Tecnologico and the Ministerio da Ciencia, Tecnologia e Inovacao, the Deutsche Forschungsgemeinschaft and the Collaborating Institutions in the Dark Energy Survey. The Collaborating Institutions are Argonne National Laboratory, the University of California at Santa Cruz, the University of Cambridge, Centro de Investigaciones Energeticas, Medioambientales y Tecnologicas-Madrid, the University of Chicago, University College London, the DES-Brazil Consortium, the University of Edinburgh, the Eidgenossische Technische Hochschule (ETH) Zurich, Fermi National Accelerator Laboratory, the University of Illinois at Urbana-Champaign, the Institut de Ciencies de l'Espai (IEEC/CSIC), the Institut de Fisica d'Altes Energies, Lawrence Berkeley National Laboratory, the Ludwig Maximilians Universitat Munchen and the associated Excellence Cluster Universe, the University of Michigan, NSF's NOIRLab, the University of Nottingham, the Ohio State University, the University of Pennsylvania, the University of Portsmouth, SLAC National Accelerator Laboratory, Stanford University, the University of Sussex, and Texas A\&M University.

BASS is a key project of the Telescope Access Program (TAP), which has been funded by the National Astronomical Observatories of China, the Chinese Academy of Sciences (the Strategic Priority Research Program "The Emergence of Cosmological Structures" Grant \# XDB09000000), and the Special Fund for Astronomy from the Ministry of Finance. The BASS is also supported by the External Cooperation Program of Chinese Academy of Sciences (Grant \# 114A11KYSB20160057), and Chinese National Natural Science Foundation (Grant \# 11433005).

The Legacy Survey team makes use of data products from the Near-Earth Object Wide-field Infrared Survey Explorer (NEOWISE), which is a project of the Jet Propulsion Laboratory/California Institute of Technology. NEOWISE is funded by the National Aeronautics and Space Administration.

The Legacy Surveys imaging of the DESI footprint is supported by the Director, Office of Science, Office of High Energy Physics of the U.S. Department of Energy under Contract No. DE-AC02-05CH1123, by the National Energy Research Scientific Computing Center, a DOE Office of Science User Facility under the same contract; and by the U.S. National Science Foundation, Division of Astronomical Sciences under Contract No. AST-0950945 to NOAO.

% Pan-STARRS

The Pan-STARRS1 Surveys (PS1) and the PS1 public science archive have been made possible through contributions by the Institute for Astronomy, the University of Hawaii, the Pan-STARRS Project Office, the Max-Planck Society and its participating institutes, the Max Planck Institute for Astronomy, Heidelberg and the Max Planck Institute for Extraterrestrial Physics, Garching, The Johns Hopkins University, Durham University, the University of Edinburgh, the Queen's University Belfast, the Harvard-Smithsonian Center for Astrophysics, the Las Cumbres Observatory Global Telescope Network Incorporated, the National Central University of Taiwan, the Space Telescope Science Institute, the National Aeronautics and Space Administration under Grant No. NNX08AR22G issued through the Planetary Science Division of the NASA Science Mission Directorate, the National Science Foundation Grant No. AST-1238877, the University of Maryland, Eotvos Lorand University (ELTE), the Los Alamos National Laboratory, and the Gordon and Betty Moore Foundation

% WSDB

This paper made use of the Whole Sky Database (wsdb) created by Sergey Koposov and maintained at the Institute of Astronomy, Cambridge by Sergey Koposov, Vasily Belokurov and Wyn Evans with financial support from the Science \& Technology Facilities Council (STFC) and the European Research Council (ERC).

% Software

This project made use of open source software including \textsc{matplotlib} \citep{matplotlib_Hunter:2007}, \textsc{numpy} \citep{numpy_harris2020array}, \textsc{scipy} \citep{scipy_2020SciPy-NMeth}, \textsc{healpy} \citep{healpyoneZonca2019, healpytwo2005ApJ...622..759G}, \textsc{dynesty} \citep{DynestyCodeSpeagle_2020, NestedSamplingSkilling2004, NestedSamplingSkilling2006, BoundingMethodFeroz_2009}, \textsc{emcee} \citep{EmceeForeman_Mackey_2013, emcee_DE_moves_Nelson_2013, emcee_snooker_DE_moves_ter_braak}, and \textsc{astropy} \citep{astropy:2013, astropy:2018}.

\section{Data Availability}

The data underlying this article were derived from sources in the public domain:

\begin{itemize}
    \item PS1 DR1: \url{https://panstarrs.stsci.edu/}
    \item DES DR1: \url{https://des.ncsa.illinois.edu/releases/dr1}
    \item DECaLS DR8: \url{https://www.legacysurvey.org/dr8/}
\end{itemize}

%%%%%%%%%%%%%%%%%%%%%%%%%%%%%%%%%%%%%%%%%%%%%%%%%%


 
\bibliographystyle{mnras}
\bibliography{references} % if your bibtex file is called example.bib

%\begin{thebibliography}{99}
%\bibitem[\protect\citetauthoryear{Author}{2012}]{Author2012}
%\bibitem[\protect\citetauthoryear{Others}{2013}]{Others2013}
%\end{thebibliography}

%%%%%%%%%%%%%%%%%%%%%%%%%%%%%%%%%%%%%%%%%%%%%%%%%%

%%%%%%%%%%%%%%%%% APPENDICES %%%%%%%%%%%%%%%%%%%%%

\appendix

\section{Rotation Matrices}
\label{sec:rot_matrix_appendix}

To transform between equatorial coordinates ($\alpha, \delta$) and the stream aligned coordinate system defined in Section \ref{sec:coordsys} ($\phi_1, \phi_2$), we use

\begin{equation}
    \begin{bmatrix}
        \cos(\phi_1) \cos(\phi_2)\\
        \sin(\phi_1) \cos(\phi_2)\\
        \sin(\phi_2)
    \end{bmatrix}
    =
    \textrm{\textbf{R}}
    \times
    \begin{bmatrix}
        \cos(\alpha) \cos(\delta)\\
        \sin(\alpha) \cos(\delta)\\
        \sin(\delta)
    \end{bmatrix}
\end{equation}

where \textbf{R} is the rotation matrix for a given stream from Table \ref{table:rot_matrix_appendix}.

\begin{table*}
    \centering
    \begin{tabular}{lccccccccc}
        \hline
        Stream      &        $R_{11}$ &        $R_{12}$ &        $R_{13}$ &        $R_{21}$ &        $R_{22}$ &        $R_{23}$ &        $R_{31}$ &        $R_{32}$ &        $R_{33}$ \\
        \hline
        Aliqa Uma   &  0.676426 &  0.472887 & -0.564647 &  0.729329 & -0.323268 &  0.602974 & -0.102606 &  0.819680 &  0.563557 \\
        ATLAS       &  0.836979 &  0.294819 & -0.461030 &  0.516168 & -0.705140 &  0.486157 &  0.181762 &  0.644871 &  0.742363 \\
        Chenab      &  0.521260 & -0.342146 & -0.781808 &  0.818214 & -0.059967 &  0.571777 & -0.242514 & -0.937732 &  0.248690 \\
        Elqui       &  0.722881 &  0.229558 & -0.651726 & -0.662398 &  0.498665 & -0.559072 &  0.196653 &  0.835845 &  0.512534 \\
        GD 1        & -0.477630 & -0.173843 &  0.861190 &  0.510845 & -0.852445 &  0.111245 &  0.714778 &  0.493068 &  0.495960 \\
        Indus       &  0.455572 & -0.165815 & -0.874620 & -0.184451 &  0.943594 & -0.274968 &  0.870880 &  0.286592 &  0.399290 \\
        Jhelum      &  0.602492 & -0.204270 & -0.771542 & -0.135321 & -0.978842 &  0.153483 & -0.786570 &  0.011933 & -0.617386 \\
        Ophiuchus   & -0.457706 & -0.881226 & -0.118092 &  0.855436 & -0.472677 &  0.211675 & -0.242353 & -0.004135 &  0.970179 \\
        Palomar 5   & -0.656057 & -0.754711 &  0.000636 &  0.609115 & -0.528995 &  0.590883 & -0.445608 &  0.388045 &  0.806751 \\
        Phoenix     &  0.614698 &  0.294787 & -0.731606 &  0.463208 &  0.615834 &  0.637328 &  0.638424 & -0.730650 &  0.242004 \\
        Triangulum  &  0.829056 &  0.359765 &  0.428060 &  0.322402 &  0.317921 & -0.891618 & -0.456863 &  0.877209 &  0.147585 \\
        Tucana III  &  0.505523 & -0.010544 & -0.862749 &  0.081198 &  0.996069 &  0.035404 &  0.858984 & -0.087951 &  0.504392 \\
        Turranburra &  0.283876 &  0.862362 & -0.419222 & -0.900340 &  0.089324 & -0.425921 & -0.329851 &  0.498351 &  0.801776 \\
        \hline
    \end{tabular}
    \caption{Rotation matrix \textbf{R} from RA and Dec to stream aligned coordinates $\phi_1$ and $\phi_2$ for each stream.  $R_{ij}$ is the element of \textbf{R} at row $i$ and column $j$. Only truncated values for $R_{ij}$ are included here.  A machine readable version of this table with full precision is included with the online version of this paper and is also publicly available on Github.}
    \label{table:rot_matrix_appendix}
\end{table*}

\section{Additional Data Files}

Two sets of supplementary files are included with the online version of this paper and are publicly available via GitHub\footnote{\label{github_repo}\url{https://github.com/jmpatric-cmu/Uniform-Modelling-Thirteen-Streams}}.  We include a separate file for each type of cubic spline (stream track ($\phi_2$), stream width ($\sigma$), distance modulus, and linear density) containing the $\phi_1$ node positions, mean node values, and $16^{th}$/$84^{th}$ percentile values for all streams (Table \ref{table:node_table}).  We also include separate files for each stream containing sample points along its stream track, in both $\phi_1$/$\phi_2$ and RA/Dec (Table \ref{table:stream_track_table}).  We only include points between the endpoints of the stream in Table \ref{table:stream_track_table}, not the full $\phi_1$ region fit by the spline.

\begin{table*}
    \centering
    \begin{tabular}{lcccc}
        \hline
        Stream      & $\phi_1$ & Mean Value & $16^{th}$ Percentile & $84^{th}$ Percentile \\
                    & [deg]    & [deg]      & [deg]                & [deg] \\
        \hline
        Aliqa Uma   & -7.5  & -0.7702  & -1.3692 & -0.0789              \\
        Aliqa Uma   & -7.0  & 0.7356   & 0.5631  & 0.9283               \\
        Aliqa Uma   & -5.5  & -0.0254  & -0.2493 & 0.2051               \\
        Aliqa Uma   & -3.5  & 0.3431   & 0.1792  & 0.5798               \\
        Aliqa Uma   & -1.5  & 0.2678   & 0.1612  & 0.3845               \\
        Aliqa Uma   & 2.5   & -0.2032  & -0.3246 & -0.1052              \\
        Atlas       & -13.0 & -0.5482  & -0.6735 & -0.4207              \\
        Atlas       & -8.9  & 0.2819   & 0.2639  & 0.2995               \\
        Atlas       & -4.8  & 0.5637   & 0.5484  & 0.5789               \\
        Atlas       & -0.7  & 0.8227   & 0.7830  & 0.8650               \\
        ...         & ...   & ...      & ...     & ... \\
        \hline
    \end{tabular}
    \caption{Stream track ($\phi_2$) cubic spline node positions and values for all streams.  We only include the first few lines of the table here.  A machine readable version of the full table for $\phi_2$, in addition to tables for the other cubic splines (stream width ($\sigma$), distance modulus, and linear density), is included with the online version of this paper and is also publicly available on Github.}
    \label{table:node_table}
\end{table*}

\begin{table*}
    \centering
    \begin{tabular}{cccc}
        \hline
        $\phi_1$    &   $\phi_2$    &   RA      &   Dec \\\relax
        [deg]       &   [deg]       &   [deg]   &   [deg] \\
        \hline
        -6.90    & 0.91     & 41.83 & -38.61 \\
        -6.81    & 1.02     & 41.84 & -38.46 \\
        -6.71    & 1.08     & 41.81 & -38.35 \\
        -6.62    & 1.11     & 41.74 & -38.27 \\
        -6.52    & 1.09     & 41.64 & -38.21 \\
        -6.43    & 1.05     & 41.52 & -38.18 \\
        -6.33    & 0.98     & 41.37 & -38.17 \\
        -6.24    & 0.89     & 41.20 & -38.17 \\
        -6.14    & 0.78     & 41.02 & -38.18 \\
        -6.05    & 0.66     & 40.83 & -38.20 \\
        ... & ... & ... & ... \\
        \hline
    \end{tabular}
    \caption{Sample points along Aliqa Uma's stream track between the two endpoints of the stream, in stream aligned coordinates $\phi_1$ and $\phi_2$ in addition to RA and Dec.  Only the first few sets of points are included here.  A machine readable version of the full table for Aliqa Uma, in addition to tables for the other streams, is included with the online version of this paper and is also publicly available on Github.}
    \label{table:stream_track_table}
\end{table*}

%%%%%%%%%%%%%%%%%%%%%%%%%%%%%%%%%%%%%%%%%%%%%%%%%%

% Don't change these lines
\bsp	% typesetting comment
\label{lastpage}
\end{document}